\documentclass[12pt]{article}
\usepackage{graphicx}
\usepackage{amsmath}
\usepackage{natbib}
\usepackage{color}
\usepackage{mathrsfs,amssymb}
\usepackage{algorithm}
\usepackage{algorithmic}
\usepackage{multirow}
\usepackage{float}
\usepackage{threeparttable}
\usepackage{color}
\usepackage{epstopdf}
\usepackage{bm,hyperref}

\usepackage{amssymb}
\usepackage{subfig}
\usepackage{enumerate,booktabs}
\usepackage{graphicx,setspace,lscape,longtable,multirow}
\usepackage{natbib,epsfig,graphicx}
\usepackage{amsmath,amsthm,amssymb,color}

\bibpunct{(}{)}{;}{a}{,}{,}

\newtheorem{theorem}{Theorem}
\newtheorem{lemma}{Lemma}
\newtheorem{prop}{Proposition}
\newtheorem{coro}{Corollary}
\def\beq{\begin{equation}}
\def\eeq{\end{equation}}
\def\beqs{\begin{equation*}}
\def\eeqs{\end{equation*}}
\def\beqr{\begin{eqnarray}}
\def\eeqr{\end{eqnarray}}
\def\beqrs{\begin{eqnarray*}}
\def\eeqrs{\end{eqnarray*}}
\def\bet{\begin{theorem}}
\def\eet{\end{theorem}}
\def\bel{\begin{lemma}}
\def\eel{\end{lemma}}
\def\bep{\begin{prop}}
\def\eep{\end{prop}}
\def\bec{\begin{coro}}
\def\eec{\end{coro}}
\def\bg{\begin{figure}[tbph]\begin{center}}
\def\eg{\end{center}\end{figure}}

\def\bc{\begin{center}}
\def\ec{\end{center}}

\def\wh{\widehat}
\def\wt{\widetilde}

\def\1{\mbox{\boldmath $1$}}

\def\mR{\mathbb{R}}
\def\mS{\mathcal{S}}
\def\mL{\mathcal L}

\def\sxx{\widehat\Sigma_{xx}}
\def\sxy{\widehat\Sigma_{xy}}

\def\Ht{\wh{\theta}}
\def\HD{\wh{\Delta}}

\def\ols{\wh\theta_{\operatorname{ols}}}

\def\argmin{\mbox{argmin}}
\def\argmax{\mbox{argmax}}

\def\wt{\widetilde}
\def\wh{\widehat}

\def\bmgd{\widehat{\theta}_{\mbox{\sc bmgd}}}

\numberwithin{equation}{section}

\begin{document}

\begin{center}
	{\bf\Large Mini-batch Gradient Descent with Buffer}\\
	\bigskip
Haobo Qi$^{1}$, Du Huang$^{2}$, Yingqiu Zhu$^{3,*}$, Danyang Huang$^4$ and Hansheng Wang$^5$\\
{\it\small 
$^1$ School of Statistics, Beijing Normal University, Beijing, China;\\
$^2$ Matpool, Timestamp Information Technology LLC, Hangzhou, China;\\
$^3$ School of Statistics, University of International Business and Economics, Beijing, China;\\
$^4$ School of Statistics, Renmin University of China, Beijing, China;\\
$^5$ Guanghua School of Management, Peking University, Beijing, China.}
\end{center}


\begin{singlespace}
\begin{abstract}
In this paper, we studied a buffered mini-batch gradient descent (BMGD) algorithm for training complex model on massive datasets. The algorithm studied here is designed for fast training on a GPU-CPU system, which contains two steps: the buffering step and the computation step. In the buffering step, a large batch of data (i.e., a buffer) are loaded from the hard drive to the graphical memory of GPU. In the computation step, a standard mini-batch gradient descent (MGD) algorithm is applied to the buffered data. Compared to traditional MGD algorithm, the proposed BMGD algorithm can be more efficient for two reasons. First, the BMGD algorithm uses the buffered data for multiple rounds of gradient update, which reduces the expensive communication cost from the hard drive to GPU memory. Second, the buffering step can be executed in parallel so that the GPU does not have to stay idle when loading new data. We first investigate the theoretical properties of BMGD algorithms under a linear regression setting. The analysis is then extended to the Polyak-\L ojasiewicz loss function class. The theoretical claims about the BMGD algorithm are numerically verified by simulation studies. The practical usefulness of the proposed method is demonstrated by three image-related real data analysis.
\end{abstract}
\end{singlespace}

\noindent {\bf KEYWORDS}: Buffered mini-batch gradient descent, Mini-batch gradient descent, Communication cost, Massive dataset, Parallel computation

\section{Introduction}

\quad\ Modern statistical analysis and machine learning problems often encounter challenges due to complex models and massive datasets. On the one hand, complex models usually involve a large number of parameters, which lead to high time cost due to {\it computation}. On the other hand, due to the dramatic improvements in data collection technology in the past few decades, massive datasets are emerging at an unprecedented speed. These datasets are often stored on hard drives, but must be processed by either CPUs and/or GPUs, which leads to significant time cost due to {\it communication} (i.e., I/O cost) when transferring data from the hard drive to CPUs and/or GPUs \citep{2017IOcost,2018IOcost}. To illustrate this concept more clearly, we consider an important benchmark dataset ImageNet2012 and the classic deep convolutional neural network AlexNet \citep{2012AlexNet} as an example. The dataset contains a total of 1,431,167 images in 1,000 classes, which requires more than 140GB space on the hard drive. The model contains more than 60 million parameters with input image size $227\times 227\times 3$. To train the models based on the dataset, a stochastic gradient descent algorithm is often used. Specifically, the data must be partitioned into a number of mini-batches, pre-loaded into the system RAM of CPU, and then loaded into the graphical memory of GPU for computation. Then, model parameter estimates are updated repeatedly. As reported by \cite{2012AlexNet}, a total of 96--120 hours are typically required needed to train the AlexNet model on the ImageNet dataset with two NVIDIA GTX 580 3GB GPUs.

To reduce the time cost due to computation, various solutions have been proposed. A promising solution is to adopt a distributed or parallel computational scheme. The main idea is to improve the overall computational efficiency by using more computing nodes. This scheme has been successfully used to deal with massive datasets for a number of statistical inference methods, including Bayesian inference \citep{doi:10.1080/01621459.2018.1429274}, generalized linear models \citep{TANG2020104567}, moment methods for high-dimensional correlated data analysis \citep{doi:10.1080/01621459.2020.1736082}, maximum quasi-likelihood estimation \citep{doi:10.1080/01621459.2020.1773832} and many others. Distributed and parallel computation can also be applied in deep learning. For example, multiple GPUs can be used to accelerate the computation process. In this case, each GPU is assigned a fixed amount of graphical memory for loading and processing input data. By using the integrated graphical memory from multiple GPUs, the size of mini-batch can be considerably enlarged. This leads to a faster speed for processing the whole sample \citep{2017mGPU}. As another effective solution, various accelerating algorithms have also been developed. For example, to reduce estimation variability, the gradient estimates from
different steps can be combined to form a more stable one \citep{tseng1998incremental,cotter2011better,lan2012optimal}. Many researchers have also studied the problem of adaptive learning rate specification \citep{duchi2011adaptive,kingma2014adam,ZHU202111}. By doing so, different learning rates can be automatically used in different dimensions of model parameter estimates. Specifically, larger learning rates should be used for directions with slower loss change (i.e., small gradient norm). In contrast, much smaller learning rates should be used for directions with faster loss change (i.e., large gradient norm). In this regard,  \cite{duchi2011adaptive} proposed AdaGrad to iteratively decrease the learning rate according to element-wised squared gradient norm. Further generalizations have been shown in RMSProp \citep{2012RMSprop}, Adadelta \citep{2012AdaDelta} and Adam \citep{kingma2014adam}. \cite{ZHU202111} proposed an improved SGD algorithm, which automatically specifies the learning rate by a method of local quadratic approximation.  Furthermore, to obtain better generalization performances with affordable time cost, various ensemble learning methods have been proposed for deep neural networks. Instead of training the model for multiple times, the idea of cyclic learning rates have been adopted to obtain multiple local estimates for further ensembling or aggregation. For example, the snapshot ensemble method of \citet{Snapshot2017}, the fast geometric ensembling method of \citet{FGE2018}, the stochastic weight averaging (SWA) method of \citet{SWA2018}.

In this work, we consider data communication for further time cost reduction. This seems to be a promising direction for two reasons. First, the data have to be loaded from a hard drive into the system RAM and GPU graphical memory before computation. Second, ample empirical evidence suggests that this step can be the major part of time cost for many typical deep learning problems \citep{2019echoing}. Ideally, one might wish to load a dataset that is as large as possible into memory. Unfortunately, this seems to be practically difficult. First, the size of the entire dataset is often much larger than the system RAM of the CPU. Furthermore, they should be passed to the GPU graphical memory for computation. Unfortunately, the size of the GPU graphical memory is even smaller. Second, a significant part of the GPU memory must be reserved to place the model. Consider for example, the VGG model of \citet{KarenSimonyan2015VeryDC} with approximately 130 million parameters. Third, another important part of the GPU graphical memory must be reserved for intermediary computation. These reasons make the effective space in GPU memory useful for raw data processing even smaller. Thus, traditional algorithms have to load datasets of much tiny sizes (i.e. mini-batches) into both the CPU and GPU memory for computation. Such a process must be repeated many times so that the entire dataset can be fully processed. This leads to the practically most popularly used mini-batch gradient descent algorithm, which leads to a significant time cost due to data communication.

The high communication cost leads to an embarrassing situation, where the GPUs compute quickly while the data loading process could be relatively slower. Thus, for a significant amount of time, the GPU is not busy with computation, but is simply waiting for the next mini-batch to be loaded in to the GPU graphical memory. Consequently, the computational power of GPUs might be largely wasted. Thus how to reconcile such an embarrassing conflict for better GPU efficiency becomes a critical problem \citep{2010NIPS,2016ICML,2019echoing}. This problem could be solved by: either (1) reducing the communication time directly by using advanced hardware technology, or (2) reducing the amount of data must be communicated by algorithmic innovation. In the past few years, significant technology progress has been achieved along the first direction. For example, NVDIA has developed the GPUDirect\footnote{\url{https://developer.nvidia.com/gpudirect}} Storage technology in 2019, which enables a direct data path between the hard drive and the GPU graphical memory. Furthermore, their Remote Direct Memory Access (RDMA\footnote{\url{https://developer.nvidia.com/blog/benchmarking-gpudirect-rdma-on-modern-server-platforms/}}) technology allows direct communications between different GPUs. However, to enforce these technologies, substantial engineering labors and financial expenses are necessarily needed. Therefore, they have not been widely adopted for common practitioners and researchers. Then, the second choice naturally becomes the primary focus of this study. Existing works have shown that it is possible to reduce communication costs based on the strategy of ``divide-and-conquer''. For example, \cite{doi:10.1080/01621459.2018.1429274} have exploited the local likelihood on distributed data blocks to provide a surrogate to the global likelihood and thus proposed a communication-efficient framework for inference problems, e.g., Bayesian inference. \cite{JMLR:v21:19-996} have proposed a framework to doubly divide data into a number of blocks and combine block-based estimations, which are vectors with relatively small sizes, to make inference for high-dimensional datasets.

Along this research direction, many novel algorithms have been proposed. For example, \cite{2010NIPS} and \cite{2016ICML} discussed parallel stochastic gradient descent (SGD), which uses local data to calculate gradients and updates separately before communicating globally. \cite{2020BA} described a data augmentation method, in which a given mini-batch of data should be repeatedly used after applying appropriate data augmentation techniques. \cite{2019echoing} proposed a data echoing method that repeatedly using the already loaded mini-batch data while waiting for another mini-batch of data to be prepared. The aforementioned studies have demonstrated that the repeated and more in-depth use of each batch of data could be helpful in practice. However, to our knowledge, the above echoing methods suffer from one common limitation. There exists little solid theoretical understanding of these training strategies. As a consequence, the theoretical properties of the algorithms remain largely unknown. Important problems such as the convergence rate of the algorithm and statistical efficiency of the resulting estimator also cannot be addressed.  Recently, \cite{yang2021basgd} have proposed a buffered asynchronous SGD method for Byzantine learning, which aims to deal with communication failure or malicious attacks within a asynchronous distributed system. They have provided rigorous theoretical treatment for their method from an optimization perspective. However, it is noteworthy that the buffering mechanism studied in their work is totally different from ours. The key difference is the definition of ``buffer''. In their work, a ``buffer'' refers to the temporally stored gradients on the master. However, in our paper, a “buffer” refers to a large subset of the whole sample that can be loaded into the system RAM of the CPU on a single machine. This makes the theoretical analysis of our work to be totally different from theirs.

To improve the communication efficiency, various methods have been proposed to fully utilize the GPU capacity under the CPU-GPU framework. They include, but are not limited to, data processing pipelines \citep{nitzberg1997collective,ma2003improving,bauer2011cudadma}, asynchronous data loading \citep{zhu2019efficient,ofeidis2022overview}, and many others. These methods share a common strategy that the data should be pre-loaded so that the idling time of GPU can be reduced and the computation capacity can be better utilized. By this strategy, we should divide a dataset into several large batches (i.e., buffers), which can be buffered into the system RAM of the CPU. However, the sizes of the buffers should be much larger than the size of a common mini-batch. Then the buffer should be further divided into many mini-batches with usual sizes, so that a standard MGD algorithm can be applied by treating the buffered data as if it were the entire dataset. The computation should be repeatedly conducted so that the GPUs can be kept busy and their computational power can be fully utilized. In the meanwhile, the next buffer of data can be prepared by the CPU in an asynchronous manner. The estimate obtained in the current buffer is then used as the initial estimate for the next buffer, which leads to fast convergence speed and decent resulting estimator. Although the buffering mechanism has been already adopted in applications of high-performance computing, the theoretical understanding remains to be insufficient so far. This is perticularly true from a statistical perspective. 

Our main contribution in this work is to provide a rigorous theoretical analysis of the BMGD algorithm from a statistical perspective with a diminishing learning rate. We show theoretically that the BMGD algorithm should converge for a broad class of loss functions under appropriate regularity conditions. The statistical properties of the resulting BMGD estimator are investigated under both linear regression models with the least squared loss and general losses from the Polyak-\L ojasiewicz (P\L) function class. It is remarkable that, we find a slower diminishing learning rate strategy is allowed for the BMGD algorithm. Therefore, a faster numerical convergence can be achieved. To our best knowledge, theoretical results of similar type have not been found in the past literature. Extensive theoretical discussions about generalized linear model and subgradient methods are also considered.

The remainder of this paper is organized as follows. Section 2 introduces the methodology of the proposed BMGD algorithm. Specifically, Section 2.1 describes the basic ideas of the BMGD algorithm and Section 2.2 introduces the model notations and the detailed algorithm. Section 3 presents the comprehensive theoretical investigations of the proposed BMGD algorithm. Specifically, Section 3.1 studies the numerical convergence properties for linear regression model and Section 3.2 considers a broader class of loss function (i.e., the P\L\ function class). Section 4 reports the results of several numerical experiments. Specifically, Section 4.1 demonstrates the theoretical claims of the BMGD algorithm via simulation experiments. Sections 4.2 - 4.4 presents three image-related real data analysis (i.e., the TT100K dataset, the CatDog dataset and the ImageNet dataset) to demonstrate its practical usefulness. Finally, Section 5 summarizes this paper with a brief discussion.

\section{Methodology}
\subsection{The basic ideas}

\begin{figure}[ht]
	\centering
	\setlength{\abovecaptionskip}{1pt}
	\includegraphics[width=1\columnwidth]{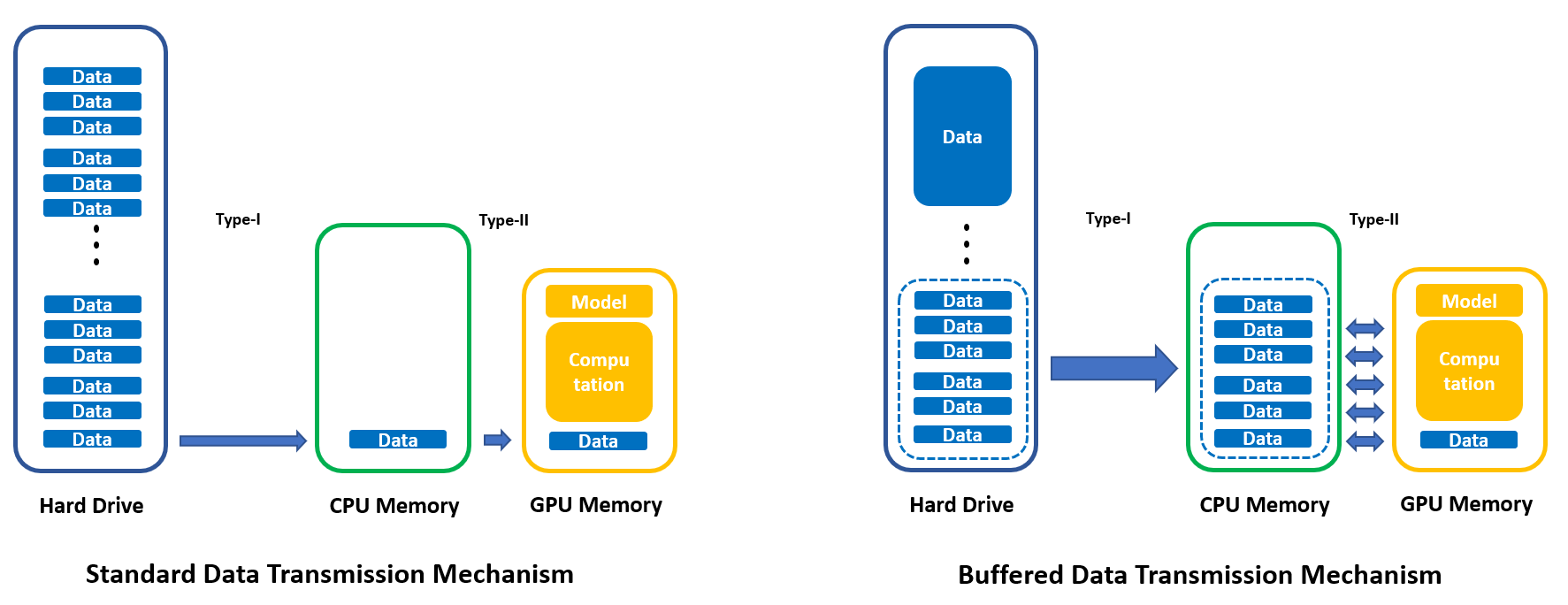}
	\caption{Illustration of the data transmission mechanism for a standard MGD algorithm (left panel) and the proposed BMGD algorithm (right panel). }
	\label{fig:buffer}
\end{figure}

Before introducing the proposed algorithm, we first present a brief description of how a standard MGD algorithm is executed on a typical CPU-GPU system. A standard CPU-GPU computation system with one single GPU (or multiple GPUs) and one shared hard drive is widely used in practice and has the following important features. First, the system's hard drive is often sufficiently large (e.g., 1 TB SSD) to store a massive dataset, while the CPU memory is much smaller (e.g., 64 GB). Unfortunately, the GPU graphical memory is typically even smaller (e.g., 16 GB). As mentioned before, the effective space for processing data in the GPU memory is even smaller. Thus, only mini-batch gradient descent methods can be practically implemented. In this study, a mini-batch refers to a sufficiently small subsample of the entire dataset. It should be copied from the hard drive to the system RAM of the CPU and then transmitted into the GPU graphical memory for gradient computing and parameter updating. By ``sufficiently small'', we mean that the size of mini-batch should be small enough so that it can be comfortably loaded in to the system RAM of the CPU, and then transmitted into the GPU graphical memory for computation. Next, the current mini-batch of data is discarded from the GPU memory, and another mini-batch should be loaded from the hard drive for the next epoch of computation. See the left panel in Figure \ref{fig:buffer} for illustration of this MGD algorithm.

Despite its practical popularity, the MGD algorithm has been criticized due to its low efficiency in computation resource usage \citep{2020BA,2019echoing}. This is mainly due to long communication times, which refer to the total time incurred while transferring data from the hard drive to the system RAM of the CPU and then to the GPU graphical memory. Compared with the GPU computation time, this communication time is large and expensive. Thus, how to reduce the communication time becomes a key issue. One possible solution is to reduce the number of communications required by the training algorithm. Thus, the data loaded in each communication step should be used more thoroughly. Traditional MGD algorithms fail to achieve this because data are discarded after only one computation step. Given that a long communication time has passed while loading those data from the hard drive to the memory, it seems natural to attempt to process those data through multiple steps before they are discarded. This concept leads to the buffered mini-batch gradient descent (BMGD) algorithm, which contains two steps that can be executed fully in parallel.
\begin{itemize}
	\item\textbf{The Buffering Step.} The objective of this step is to buffer a relatively large batch of data from the hard drive to the CPU memory. As we discussed before, this step requires expensive communication costs. Therefore, we wish to load as much data as possible in parallel while the GPU is fully engaged in computing. Also, we wish to decrease the number of repeated and possibly redundant loadings for the same batch of data to be as small as possible.
	
	\item\textbf{The Computation Step.} The objective of this step is to apply a standard MGD algorithm to the buffered data. Specifically, the buffered data in the CPU memory should be further divided into many mini-batches. They are then sequentially processed into the graphical memory of GPU for gradient computation and parameter updating. By treating the buffered data in the GPU memory as if it were the entire dataset, a standard MGD algorithm can be performed repeatedly for many epochs until the next buffer of data are prepared.
\end{itemize}
See the right panel in Figure \ref{fig:buffer} for a better illustration. As we have discussed previously, for a standard MGD algorithm, the GPU has to spend a lot of time waiting for the CPU to prepare a sequence of mini-batch datasets. This delay leads to high communication time and thus a low efficiency in GPU usage. However, this conflict could be naturally avoided in the proposed BMGD algorithm because the buffering and computation steps can be executed in parallel. Thus, we leave the CPU a sufficient amount of time for data buffering and allow the GPU to perform more in-depth data calculations, which leads to more accurate parameter updating and estimation. For an integrated system with multiple GPUs, those GPUs can be used to process a given mini-batch in a fully parallel way. By doing so, the computation power of a multiple-GPU system can be fully utilized.

\begin{figure}[ht]
	\centering
	\setlength{\abovecaptionskip}{1pt}
	\includegraphics[width=1\columnwidth]{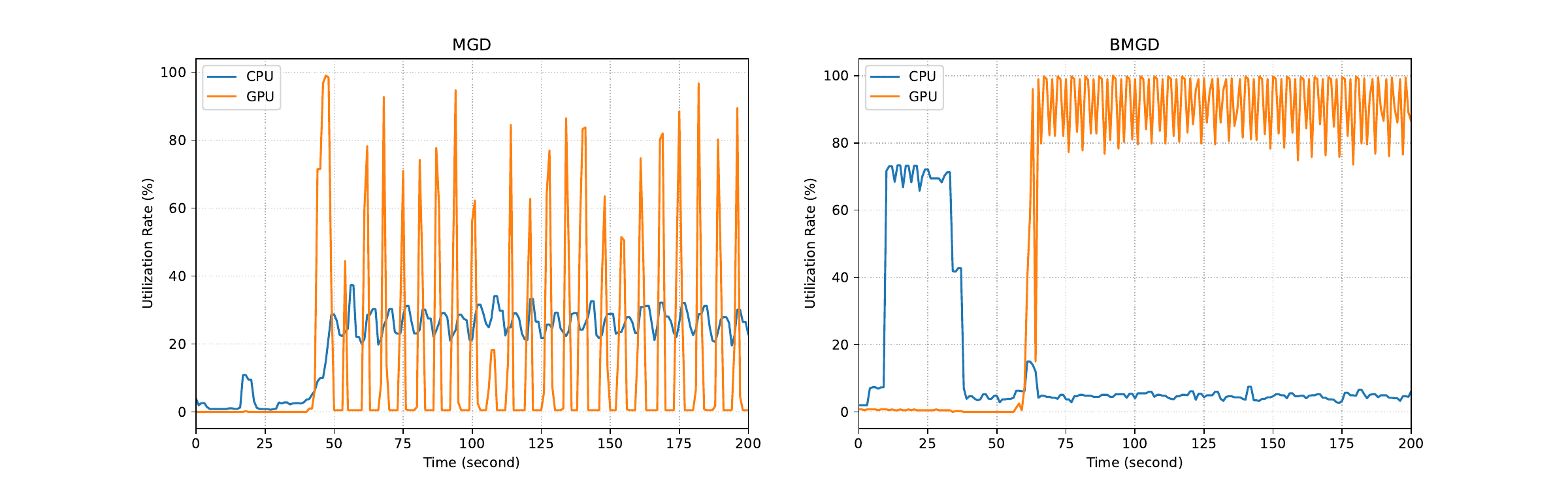}
	\caption{Graphical illustration of the CPU and GPU utilization efficiency for the MGD algorithm and the proposed BMGD algorithm. The left panel presents the MGD results and the right panel presents the BMGD results.}
	\label{fig:utilization}
\end{figure}

Compared with the standard MGD algorithm, the proposed BMGD method exhibits many advantages but also faces some new challenges. First, the BMGD algorithm uses GPU resources more efficiently. For illustration purpose, we consider a simple numerical experiment, which trains the ResNet50 model \citep{He2016Deep} on the benchmark dataset ImageNet2012 via the standard MGD algorithm and the BMGD algorithm. The input size is set to be $224\times 224 \times 3$ and the batchsize is set to be 128. The utilization rates of CPU and GPU are monitored for both two algorithms. The results of the first 200 second training are presented in Figure \ref{fig:utilization}. By Figure \ref{fig:utilization}, we find that the MGD algorithm suffers from GPU idling problem seriously during the training while the GPU utilization for BMGD algorithm remains a much higher level. Second, more in-depth calculations of each buffered data might lead to faster statistical convergence. However, the new challenge comes from more in-depth computation forcing the estimator to converge towards the local optimizer of the buffered dataset instead of the global one. If each buffered dataset is randomly generated, we should expect that these local optimizers should also be consistent estimators for the population parameter under appropriate regularity conditions. However, they should be statistically less efficient than the global one. To fix this problem, we find that a small learning rate should be used and coordinated with epoch number on each buffer. We present a detailed bound for the resulting estimator of its statistical accuracy, and associated technical details are discussed in the next subsections.

\subsection{Buffered mini-batch gradient descent}
\par Before the theoretical discussion, we first introduce some basic notations. We let $\mS = \{1,2,..,N\}$ be the index set of the entire sample, which should be placed on the hard drive. We also let $(X_i, Y_i)$ be the observation collected from the $i$-th subject, where $Y_i\in\mathbb{R}^1$ is the response of interest and $X_i=(X_{i1},X_{i2},\cdots,X_{ip})^\top\in\mR^p$ is the associated predictor. We assume that $(X_i, Y_i)$ for different subject $i$s are independently and identically distributed. To address the conditional relationship between $Y_i$ and $X_i$, a statistical model (e.g., a generalized linear model) is often assumed, which leads to a loss function evaluated at each sample $i$ as $\ell(X_i,Y_i;\theta)$, where $\theta\in\mathbb{R}^{q}$ denotes the model parameter of interest and $\theta^*$ denotes the true parameter. We allow parameter dimension $q$ and predictor dimension $p$ to be different. For most regression problems, we should have the coefficient $\theta$ sharing the same dimension as the predictor $X$. If this happens, we should have $p=q$. Then, the global loss function is defined as $\mL(\theta)=N^{-1}\sum^N_{i=1} \ell(X_i,Y_i;\theta)$.

A natural estimator of $\theta^*$ can be obtained by minimizing the global loss function as $\widehat{\theta} = \operatorname{\argmin}_\theta\mL(\theta)$, which is referred to as the global estimator. For example, if $\ell(X_i,Y_i;\theta)$ denotes the negative log-likelihood function of $(X_i, Y_i)$, then $\widehat{\theta}$ becomes the maximum likelihood estimator (MLE). If $\widehat{\theta}$ is the MLE and appropriate regularity conditions can be assumed, we should have $\sqrt{N}(\widehat{\theta}-\theta^*)\to^{d}N(0, \Sigma^{-1})$, where $\Sigma = E\{\dot{\ell}(X_i,Y_i; \theta^*)\dot{\ell}^\top(X_i,Y_i; \theta^*)\}= -E\{\ddot{\ell}(X_i,Y_i; \theta^*)\}$ is the Fisher information matrix. In this study, $\dot{\ell}(\cdot; \theta)$ and $\ddot{\ell}(\cdot; \theta)$ represent the first- and second-order derivatives of $\ell(\cdot; \theta)$ with respect to $\theta$, respectively. Unfortunately, such a natural estimator is difficult to compute in practice with massive datasets because the entire dataset cannot be loaded into memory as a whole.

\begin{figure}[ht]
	\centering
	\setlength{\abovecaptionskip}{1pt}
	\includegraphics[width=1\columnwidth]{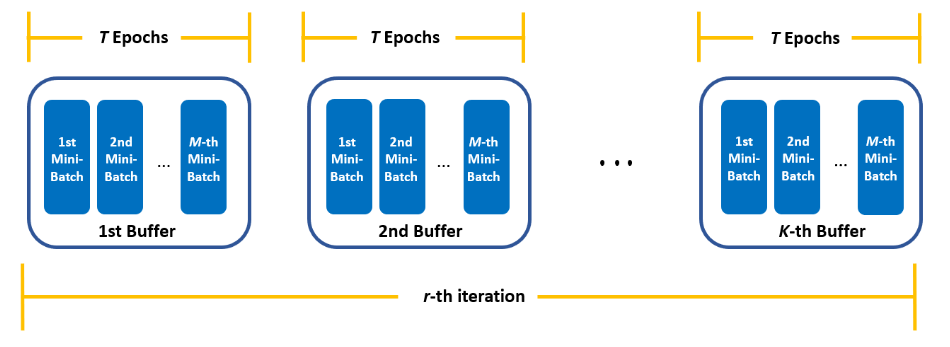}
	\caption{Illustration of the definition of {\it epoch} and {\it iteration}. }
	\label{fig:iteration}
\end{figure}

Then, we introduce some notations for the proposed BMGD algorithm. First, in the buffering step, we divide the entire sample into $K$ nonoverlapping subsamples, which is the large batch mentioned above. For simplicity, we refer to each subsample as a {\it buffer} and collect its indices by $\mS_{r,k}$ ($1\leq r\leq R, 1\leq k\leq K$). For all $1\leq r\leq R$, we have $\mS=\bigcup_k \mS_{r,k}$ and $\mS_{r,k_1}\cap \mS_{r, k_2}= \emptyset$ for $k_1\neq k_2$. Let $N_{r,k}=|\mS_{r,k}|$ be the size of the $k$-th buffer in the $r$-th iteration. For convenience, we assume that $N_{r,k} = N_0$ for every $1\leq k\leq K$. We then have $N = \sum_{k} N_{r,k}=K N_0$. The size of the buffer (i.e., $N_0$) should also be carefully designed. On one side, it should be small enough so that it can be easily loaded into the system RAM of the CPU. On the other side, it should also be large enough so that a sufficiently accurate local estimator can be produced by each buffer. Second, in the computation step, we decompose each buffer into $M$ {\it mini-batches}. We let $\mS_{r,k}^{(t,m)}$ be the index set of the sample calculated on the $m$-th mini-batch in the $t$-th epoch for the $k$-th buffer in the $r$-th iteration. We also assume that the size of the mini-batch is given by $n=|\mS_{r,k}^{(t,m)}|$ for every $r$, $k$, $t$ and $m$. We then should have $N_0=Mn$ and $N=KMn$.
For a given buffer (e.g., $\mS_{r,k}$), we apply a standard MGD algorithm by treating this buffer as if it were the entire sample. Thus, the buffer $\mS_{r,k}$ should then be processed from the first mini-batch $\mS_{r,k}^{(1,1)}$ until the last one $\mS_{r,k}^{(1,M)}$. We call one complete round of this mini-batch computation process for one particular buffer an {\it epoch}. This procedure is then repeated for a total of $T$ times. See Figure \ref{fig:iteration} for a graphical illustration.
Next, we describe the proposed BMGD algorithm in detail. We let $\widehat\theta^{(t,m)}_{r,k}$ be the estimator obtained in the $r$-th iteration on the $k$-th buffer and in the $t$-th epoch computed on the $m$-th mini-batch. Here, $1\leq t\leq T$ and $1\leq r \leq R$ for some prespecified number of buffer epochs $T$ and iterations $R$, respectively. First, in one single iteration, all buffers should be sequentially processed from $\mS_{r,1}$ to $\mS_{r,K}$. Then, for each buffer $\mS_{r,k}$, in the $t$-th epoch, we should update parameter estimates based on the following updating rule:
\beqr
\label{eq:1}
&\widehat{\theta}_{r,k}^{(t,1)}=\widehat{\theta}_{r,k}^{(t-1,M)}-\alpha\dot{\mL}_{r,k}^{(t,1)}\left(\widehat{\theta}_{r,k}^{(t-1,M)}\right),\nonumber\\
\widehat{\theta}_{r,k}^{(t,m)}&=\widehat{\theta}_{r,k}^{(t,m-1)}-\alpha\dot{\mL}_{r,k}^{(t,m)}\left(\widehat{\theta}_{r,k}^{(t,m-1)}\right) \mbox{\ \ for\ \ } 2\leq m \leq M,
\eeqr
where $\alpha>0$ is the learning rate, and $\dot\mL_{r,k}^{(t,m)}(\theta)=n^{-1}\sum_{i\in\mS^{(t,m)}_{r,k}}\dot\ell(X_i,Y_i;\theta)$ is the gradient computed on the sample $\mS^{(t,m)}_{r,k}$. Then, the MGD algorithm (\ref{eq:1}) should be fully processed for a total of $T$ epochs, which leads to the final estimator $\widehat{\theta}_{r,k}^{(T,M)}$ on the $k$-th buffer in the $r$-th iteration.
It is then treated as the initial estimator for the MGD algorithm on the next buffer. Once the training processes have finished on all $K$ buffers in a single iteration, the final estimator on the last buffer is then treated as the initial estimator for the next iteration i.e. $\widehat{\theta}_{r+1,1}^{(0)} =\widehat{\theta}_{r,K}^{(T,M)}$. After a total of $R$ round iterations, we obtain the final estimator, which is defined as the BMGD estimator i.e. $\bmgd = \widehat{\theta}_{R,K}^{(T,M)}$. This process ends the proposed BMGD algorithm, and the entire procedure is also given in Algorithm \ref{alg1}.

\begin{algorithm}[ht]
	\caption{Buffered mini-batch gradient descent (BMGD)} \label{alg1}
	\begin{algorithmic}[1]
		\REQUIRE Initial value $\widehat{\theta}^{(0)}_{1,1}$, Learning rate $\alpha$, Buffer number $K$, Mini-batch number $M$\\
		\ \ \ \ \ \ Buffer epoch number $T$, Iteration number $R$.
		\FOR{$r = 1,2,...,R$}
        \STATE Generate random partition of the whole sample as $\{\mS_{r,1}, ..., \mS_{r,K}\}$.
        \STATE Initial CPU to start preparing buffered data.
        \STATE  Set initial value as $\widehat{\theta}^{(0)}_{r,1} = \widehat{\theta}^{(0)}_{1,1}$ and $\widehat{\theta}^{(0)}_{r,1} = \widehat{\theta}^{(T,M)}_{r-1,K}$ for $r>1$.
		\FOR{$k = 1,2,...,K$}
		\FOR{$t = 1,2,...,T$}
        \STATE Generate random partition of the buffer $\mS_{r,k}$ as $\left\{\mS^{(t,1)}_{r,k}, ..., \mS^{(t,M)}_{r,K}\right\}$.
		\FOR{$m = 1,2,...,M$}
		\STATE  Update parameter in the $m$-th mini-batch as
		\beqrs
        \widehat{\theta}_{r,k}^{(t,m)}&=&\widehat{\theta}_{r,k}^{(t,m-1)}-\alpha\dot\mL_{r,k}^{(t,m)}\left(\widehat{\theta}_{r,k}^{(t,m-1)}\right)
		\eeqrs
		\ENDFOR
		\ENDFOR
		\ENDFOR
		\ENDFOR
		\ENSURE BMGD estimator $\bmgd = \widehat{\theta}_{R,K}^{(T,M)}$
	\end{algorithmic}
\end{algorithm}

Next, we consider the communication cost for the BMGD algorithm. Denote the communication cost from the hard drive to the CPU memory as the Type-I cost and the communication cost from the CPU memory to the GPU graphical memory as the Type-II cost. Assume the number of buffer is $K$, the number of mini-batch in each buffer is $M$, the number of buffer epoch is $T$, and the number of iteration is $R$. This leads to a total of $(KM)$ mini-batches and $(RTKM)$ gradient updates for the whole sample data. Further assume that the Type-I cost and Type-II cost for a single mini-batch data are $C_1$ and $C_2$, respectively. Then the communication cost for the standard MGD algorithm is about $RTKM(C_1 + C_2)$, since each mini-batch of the data needs to be transmitted from the hard drive to the GPU graphical memory and then discarded after updating. However, the communication cost for the BMGD algorithm is only about $(RKMC_1 + RTKMC_2)$. This is because once the buffered data are loaded into the CPU RAM, they are then repeatedly used for a total of $T$ epochs rather than directly discarded. Consequently, the communication cost of the BMGD algorithm can be much smaller than that of the MGD algortihm when the type-I cost $C_1$ and number of buffer epoch $T$ are large.

\section{Theoretical Results}
\subsection{Convergence analysis for linear regression model}

We now investigate the theoretical properties of the proposed BMGD estimator. To start with, we consider the
classical linear regression model. Specifically, we consider a standard linear regression model as $Y_i = X^\top_i \theta_0 + \varepsilon_i$, where $\varepsilon_i$s are independently and identically distributed error terms with mean $0$ and variance $\sigma^2$. Here, $\theta_0\in\mathbb{R}^{p}$ denotes the true value of the regression coefficient to be estimated. Typically, the most commonly used global loss function for estimating $\theta_0$ is the squared loss $\mL(\theta) = N^{-1}\sum^N_{i=1}(Y_i -X^\top_i \theta)^2$. As a result, the global minimizer $\widehat{\theta}$ becomes the ordinary least square (OLS) estimator $\ols = \sxx^{-1}\sxy$, where $\sxx = N^{-1}\sum^N_{i=1}X_iX^\top_i$ and $\sxy = N^{-1}\sum^N_{i=1}X_iY_i$. We define $\Sigma_{xx} = E(X_iX^\top_i)$; then, classic statistical theory asserts that $\ols$ is asymptotically normal with $\sqrt{N}(\ols - \theta_0)\xrightarrow{d}N(0, \sigma^2\Sigma^{-1}_{xx})$ under some regularity conditions.
This value
becomes an important benchmark estimator for comparing and understanding the asymptotic behavior of the BMGD estimator.

\subsubsection{Linear system}
For simplicity purpose, we start with the assumption that $\mS^{(t,m)}_{r,k} = \mS^{(k,m)}$ for all $1\leq r\leq R$ and $1\leq t\leq T$. By doing so, we assume the partition for buffers and mini-batches to be fixed throughout the training procedure. Under the above linear regression setup, we can express the gradient and Hessian matrix of the least square
loss explicitly. Thus, the updating formula (\ref{eq:1}) can be rewritten as:
\beqr
\label{eq:2}
\widehat{\theta}_{r,k}^{(t,1)}&=&\HD^{(k,1)}\widehat{\theta}_{r,k}^{(t-1,M)} + \alpha\sxy^{(k,1)},\nonumber\\
\widehat{\theta}_{r,k}^{(t,m)}&=&\HD^{(k,m)}\widehat{\theta}_{r,k}^{(t,m-1)}+\alpha\sxy^{(k,m)} \mbox{\ \ for\ \ } 2\leq m \leq M,
\eeqr
where the terms are $\HD^{(k,m)} = (I_p-\alpha\sxx^{(k,m)})$, $\sxx^{(k,m)} = n^{-1}\sum_{i\in\mS^{(k,m)}}X_iX^\top_i$, and $\sxy^{(k,m)} = n^{-1}\sum_{i\in\mS^{(k,m)}}X_iY_i$. By iteratively applying the updating formula (\ref{eq:2}), we can further obtain that the updating formula between $(k-1)$-th and $k$-th buffers:
\beqr
\label{eq:3}
\Ht^{(T,M)}_{r,k} = \widehat{C}^{(k)}\Ht^{(T,M)}_{r,k-1} + \alpha\widehat{D}^{(k)},
\eeqr
where $\widehat{C}^{(k)} = (\widehat{A}^{(k)})^T $, $\widehat{A}^{(k)} = (\prod^M_{m=1}\HD^{(k,m)})$, $\widehat{D}^{(k)} = \{ I_p + \cdots + (\widehat{A}^{(k)})^{T-1}\} \widehat{B}^{(k)}$, $\widehat{B}^{(k)} = \sum^M_{m=1}(\prod^M_{s=m+1}\HD^{(k,s)}) \sxy^{(k,m)}$ for $1\leq k\leq K$, and $I_p\in \mR^{p\times p}$ is an identity matrix. Next, to analyze the stable solution in an iteration, we stack the estimators from all the $K$ buffers together and obtain the estimator for the BMGD algorithm in vector form. Specifically, we let $\Ht_r^{*} = (\Ht^{(T,M)\top}_{r,1},\Ht^{(T,M)\top}_{r,2},...,\Ht^{(T,M)\top}_{r,K})^\top \in \mathbb{R}^{Kp}$, and $\widehat{D}^{*} = (\widehat{D}^{(1)\top},\widehat{D}^{(2)\top},...,\widehat{D}^{(K)\top})^\top \in \mathbb{R}^{Kp}$. We also define a block matrix $\widehat{C}^{*}$ as:
\beqrs
\widehat{C}^{*} =
\left(
\begin{array}{ccccc}
	\bm{0} & \bm{0} & \cdots & \bm{0} &\widehat{C}^{(1)}\\
	\widehat{C}^{(2)} & \bm{0} & \cdots & \bm{0} & \bm{0}\\
	\vdots\\
	\bm{0} & \bm{0} & \cdots & \bm{0} & \bm{0}\\
	\bm{0} & \bm{0} & \cdots & \widehat{C}^{(K)} & \bm{0}\\
\end{array}
\right)\in \mathbb{R}^{Kp\times Kp}.
\eeqrs
Then, the updating formula (\ref{eq:3}) for the $r$-th iteration can be rewritten as:
\beq\label{eq:4}
\Ht_r^{*} = \widehat{C}^{*}\Ht_{r-1}^{*} + \alpha \widehat{D}^{*}.
\eeq
We find that this updating formula naturally forms a linear system. This interesting representation presents a simple description of the BMGD algorithm and provides a useful mathematical tool to investigate the theoretical properties of the BMGD algorithm and the resulting estimator.

\subsubsection{Stable solution and numerical convergence}

We now investigate the numerical convergence of the BMGD algorithm with the help of the linear system given in (\ref{eq:4}). According to the linear system, we can define a stable solution as $\widehat{\theta}^* = \alpha \widehat{\Omega}^{-1}\widehat{D}^*$, where $\widehat{\Omega} = I_{Kp} - \widehat{C}^{*}$. If $\widehat{\Omega}$ is invertible, the stable solution $\widehat{\theta}^*$ should exist and satisfy (\ref{eq:4}). Next, supposing that a stable solution exists, we further need to discuss whether the BMGD algorithm converges numerically to the stable solution $\widehat{\theta}^*$. If we subtract the stable solution on both sides of the linear system, we obtain $\Ht_r^{*}-\widehat{\theta}^* = \widehat{C}^{*}\left(\Ht_{r-1}^{*}-\widehat{\theta}^*\right)$. By iteratively applying this, we then have $\Ht_r^{*}-\widehat{\theta}^* = \left(\widehat{C}^{*}\right)^{r}\left(\Ht_{0}^{*}-\widehat{\theta}^*\right)$. Then, it suffices to study the numerical properties of matrix $\widehat{C}^{*}$. We summarize the key theoretical results in Theorem \ref{thm1}.
\bet[Numerical Convergence]\label{thm1}
 Assume that $\Sigma_{xx}$ is a positive definite matrix with eigenvalues $\lambda_1\geq \lambda_2 \geq \cdots \geq \lambda_p>0$. Let $\Ht^*_{0}\in\mathbb{R}^{Kp}$ be the initial estimator and $\rho = \max\{|1 - \alpha \lambda_p|^{TM}, |1-\alpha \lambda_1|^{TM}\}$, if $0<\alpha<2\lambda^{-1}_1$ then as $N\to \infty$ we have: \\
(a) $\widehat{\theta}^{*} = \alpha \widehat{\Omega}^{-1}\widehat{D}^*$ is well-defined with probability tending to one; \\
(b) $\rho <1$ and
\[
\|\Ht^*_{r} - \widehat{\theta}^{*}\|\leq \rho^{r_0}_0\wt{\rho}^{r-r_0} \|\Ht^*_{0} - \widehat{\theta}^{*}\| \text{\ \ \ \ for\ } r>r_0,
\]
holds with probability tending to one for some constants $r_0<+\infty$, $\rho<\rho_0<+\infty$ and $\rho\leq \wt{\rho}<1$.
\eet

The detailed proof can be found in Appendix A. According to Theorem \ref{thm1}, we know that under certain condition the stable solution does exist and the BMGD algorithm converges numerically to the stable solution as long as learning rate $\alpha$ is set appropriately. Furthermore, when iteration number $r\to \infty$ we find that the BMGD algorithm achieves linear convergence rate. Although the theorem results are similar to the existing results of stochastic gradient descent, there are two critical differences. First, the BMGD method studied here is a fixed and random partition based mini-batch method. Once the whole sample is partitioned into non-overlapping mini-batches, they should be periodically and repeatedly used across different epochs and iterations. As a consequence, the resulting mini-batches in the BMGD algorithm are dependent with each other. This violates the martingale-difference assumption of the gradient noise used in the existing literature \citep{Polyak1992, SA2011, Chen2020}. Second, compared with the standard mini-batch gradient descent algorithm, the BMGD method allows multiple epoch updates for one single buffered data. This further differentiates the theoretical analysis from the classical mini-batch gradient descent algorithm, where each mini-batch is only used once in one single epoch.

\subsubsection{Asymptotic properties of the stable solution}

\par According to the results in Section 2.4, we know that the BMGD algorithm should converge to the stable solution $\widehat{\theta}^{*}$ with a high probability under appropriate regularity conditions. We now investigate the statistical properties of the stable solution. More specifically, we compare the proposed method's statistical efficiency with the benchmark estimator (i.e. the global OLS estimator $\ols$). We thus have the following theorem.


\bet[Stable Solution]\label{thm2}
Assuming the assumptions in Theorem \ref{thm1} hold and $\alpha T$ is sufficiently small, then there exists constant $c>0$ such that the stable solution $\widehat{\theta}^{*}$ satisfies $\big\|\widehat{\theta}^{*} - \bm{1}_K\otimes \ols\big\| \leq c\alpha T$ holds with probability tending to one.
\eet
\noindent
By Theorem \ref{thm2}, we find that the distance between the stable solution and the OLS estimator is primarily linearly bounded by $\alpha T$. As a result, if $\alpha T\to 0$, we should expect that stable solution $\widehat{\theta}^*$ converges to the OLS estimator. Also, if $\alpha T = O_p(N^{-1/2})$, then $\widehat{\theta}^*$ is $\sqrt{N}$-consistent. If $\alpha T = o_p(N^{-1/2})$, $\widehat{\theta}^*$ should have the same asymptotic efficiency as the global optimal estimator $\ols$. Thus, we should have $\alpha = o_p(N^{-1/2}T^{-1})$, which suggests that for a particular sample size $N$, a larger number of epochs $T$ should be accompanied by a smaller learning rate $\alpha$. Otherwise, the resulting BMGD estimator might not be statistically efficient. Such an interesting finding is not surprising. If we assume $T\to\infty$, we then should expect the MGD estimator computed on each buffer to be close to the local OLS estimators, which is markedly less efficient than the global estimator. A smaller learning rate also leads to better statistical efficiency but also leads to slower numerical convergence speed (see Theorem \ref{thm1}). This might be a price, which has to be paid for balancing statistical efficiency and convergence speed.

\subsection{Convergence analysis for general loss functions}

\subsubsection{Constant learning rates}
In this subsection, we try to extend the theoretical analysis of the BMGD algorithm to a more general loss function. Since finding the global minimizer of general loss is a long-standing difficult problem, following the theoretical framework of \citet{Nguyen2020AUC} and \citet{Ahn2020SGDWS}, we focus on the suboptimality of objective function value, $(\mL(\theta) - \mL^*)$. Here $\mL^* = \min_{\theta\in\Theta}\mL(\theta)$. Specifically, we consider here the Polyak-\L ojasiewicz (P\L) function class, which contains those loss function $\mL(\theta)$ satisfying $\mu$-P\L condition as $\|\dot{\mL}(\theta)\|^2\geq 2\mu (\mL(\theta) - \mL^*)$ for any $\theta\in\Theta$. The P\L\ class is a broad class of loss functions including many typically encountered convex loss function (e.g., the negative log-likelihood function of a generalized linear model) as special cases. In the meanwhile, the P\L-function class also allows the loss function to be nonconvex to some extend. Therefore it has been widely used in the optimization literature \citep{HaoChen2019RandomSB, Nguyen2020AUC, Ahn2020SGDWS}. We next assume the following technical conditions.
\begin{itemize}
	\item[(A1)]({\sc Parameter Space}) The parameter space $\Theta$ is compact and the loss function satisfies that $\mL^* = \min_{\theta\in\Theta}\mL(\theta)>-\infty$.
	\item[(A2)]({\sc Bounded Gradient}) For any $(x,y) \in \mathcal{D}$ and $\theta \in \Theta$, there exists a positive constant $G$ such that $\|\dot{\ell}(x,y; \theta)\|_2 \leq G$.
	\item[(A3)] ({\sc $L$-smoothness}) There exists a positive constant $L$ such that $\mL_{r,k}^{(t,m)}(\theta_2) \leq \mL_{r,k}^{(t,m)}(\theta_1) + \dot{\mL}_{r,k}^{(t,m)}(\theta_1)^\top (\theta_2 - \theta_1) + L^2/2\|\theta_1 - \theta_2\|^2$ hold for any $1\leq m \leq M$, $1\leq t\leq T$, $1\leq k\leq K$, $1\leq r\leq R$, and $\theta_1, \theta_2\in \Theta$. 
	\item[(A4)] ({\sc $\mu$-Polyak-\L ojasiewicz Condition}) There exists a positive constant $\mu$ such that $\|\dot{\mL}(\theta)\|^2\geq 2\mu (\mL(\theta) - \mL^*)$ for any $\theta\in\Theta$.
\end{itemize}
Assumption (A1) specifies a compact parameter space. Assumption (A2) requires the gradients involved to be bounded. Both the assumptions have been widely used in the optimization literature; see for example,  \citet{2019Understanding}, \citet{Nguyen2020AUC}, and \citet{Ahn2020SGDWS}.  Assumption (A3) assumes the loss function should be first-order continuously differentiable and the Hessian norm of the loss function should be uniformly bounded by a positive constant $L$. Assumption (A4) assumes the global loss function to satisfy the P\L-condition, which is a nonconvex relaxiation of the strongly convex property. Both of (A3) and (A4) are standard assumptions used in convergence analysis of gradient based methods; see for example \citet{2019Understanding}, \citet{pmlr-v119-assran20a}, and \citet{Ahn2020SGDWS}. Following these literatures, we define $\kappa = L/\mu$. With the above assumptions, we then obtain the following theorem.

\bet\label{thm3}
Assume that assumptions (A1) - (A4) hold. Further assume an arbitrary iteration number $R$ satisfying $R\geq 12\kappa \log\{(MTK)^{1/2}R\}$ and a fixed learning rate as $\alpha = 2\log\{(MTK)^{1/2}R\}/(\mu MTK R)$. Conditional on the whole sample and assume an arbitrary constant $0<\delta<1$, then with probability at least $1-\delta$ there exists a constant $c = O(L^2G^2/\mu^3)$ such that
\beqrs
 \min_{1\leq r\leq R}\mL(\wh\theta_{r,K}^{(T,M)}) - \mL^* \leq \frac{\left(\mL(\wh\theta^{(0)}) - \mL^*\right)}{MTK R^2} + \frac{c\log^2\{(MTK)^{1/2}R\}}{R^2}\log \left(\frac{KR}{\delta}\right),
\eeqrs
\eet

The detailed proof of Theorem 3 can be found in Appendix A.4. By Theorem 3, we find that the convergence upper bound of the proposed BMGD algorithm is mainly determined by two parts. The first part is due to the diminishing effect of initial value $\wh\theta^{(0)}$. It is similar to the lower bound $O\{1/(KR^2)\}$ of \citet{Ahn2020SGDWS} but with an extra factor $1/(TM)$. This extra factor is due to the multiple MGD updates conducted on each buffer. This leads to an accelerated diminishing effect. The second part is the convergence rate. With a fixed learning rate $\alpha = 2\log\{(MTK)^{1/2}R\}/(\mu MTK R)$, the convergence rate of our algorithm is approximately of order $O[\log^2\{(MTK)^{1/2}R\}/(KR^2)]$. This is also similar to the results of \citet{Ahn2020SGDWS} but with an extra factor $\log^2(MT)$. This suggests that increasing the number of mini-batches $M$ and the number of epoch $T$ on each buffer would enlarge the upper bound of convergence rate and thus the worse generalization performance. This seems to be a price we have to pay for faster convergence of the first term (i.e., the diminishing effect). To partially fix the problem, the learning rate $\alpha$ should be carefully set according to other tuning parameters ( i.e. $M, T, K, n$ and $R$). Take the mini-batch size $n$ as an example. Note that $n = N/(MK)$, the learning rate $\alpha$ for obtaining optimal convergence rate should be estimated as $\alpha = c_{\alpha}n\log\{(NT/n)^{1/2}R\}/(TN R)$ for some constant $c_{\alpha}>0$. One simple and effective solution is to set the learning rate $\alpha$ proportional to $1/(MTK)$. Since the benefits obtained by the accelerated effect is larger than the price paid for the convergence rate, the overall convergence performance is improved.

\subsubsection{Diminishing learning rates}

The previous section focus on the convergence analysis of the BMGD algorithm with a constant learning rate for the P\L\ function class. To obtain a better error upper bound, a tiny learning rate is often needed for essentially any type of mini-batch (or stochastic) gradient descent algorithms. However, too small a learning rate could lead to a painfully slow numerical convergence speed. A more common practice is to use a relatively larger learning rate in the early stage of the training process and then reduce the learning rate following some strategy \citep{robbins1951stochastic, ZHU202111}. For convenience, we refer to this as a diminishing learning rate strategy. Then how to appropriately specify the diminishing learning rates for efficient numerical convergence of the proposed BMGD algorithm becomes a problem of interest. Specifically, let $\alpha_r$ and $T_r$ be the learning rate and number of buffer epoch in the $r$-th iteration, respectively. Based on the previous analysis, we then have the following theorem.


\bet\label{thm4}
Assume that assumptions (A1) - (A4) hold and $\alpha_1T_1 \leq 1/(6KM)$. Conditional on the whole sample and assume an arbitrary constant $0<\delta<1$, then there exists a constant $c = O(L^2G^2/\mu^3)$ such that with probability at least $1-\delta$, we have
\beqrs\label{eq:3.4}
\min_{1\leq r\leq R}\mL(\wh\theta_{R,K}^{(T,M)}) - \mL^* &\leq&\frac{\left(\mL(\wh\theta^{(0)}) - \mL^*\right)}{\exp\Big\{\mu KM\big(\sum^R_{r=1}\alpha_rT_r\big)\Big\}}\\
&&+c(MK)^3\log\left(\frac{KR}{\delta}\right)\sum^R_{r=1}\frac{(\alpha_rT_r)^3}{\exp\Big\{\mu KM\big(\sum^R_{s=r+1}\alpha_s T_s\big)\Big\}}.
\eeqrs
\eet

The detailed proof of Theorem \ref{thm4} can be found in Appendix A.5. By Theorem \ref{thm4}, we can draw the following conclusions. Firstly, compared with traditional theoretical results on diminishing learning rates in the literature \citep{robbins1951stochastic, nemirovski2009robust, gitman2019understanding}, Theorem \ref{thm4} is more flexible since it involves both learning rates $\alpha_r$ and number of buffer epoch $T_r$. If $T_r$ is fixed as $T_r = T_0$ for some positive integer $T_0$, Theorem \ref{thm4} reduces to a conclusion about diminishing learning rates only. Secondly, we find that the excessive loss can be decomposed into two parts. The first part represents the diminishing effect of the initial value. In this regard, the sequence $\{\alpha_r T_r\}^{\infty}_{r=1}$ should not vanish too fast in the sense that $\sum^{\infty}_{r=1}\alpha_r T_r= \infty$. Otherwise, the loss difference due to the first part might not converge to zero. The second part describes the algorithmic convergence rate of the BMGD algorithm with a diminishing sequence $\{\alpha_rT_r\}^{\infty}_{r=1}$. A pair of sufficient conditions for the second part to diminish are (i) $\sum^{\infty}_{r=1}\alpha_r T_r= \infty$ with $\alpha_R T_R = o\{1/(\log R)\}$, and (ii) $\sum^{\infty}_{r=1}(\alpha_r T_r)^3 < +\infty$ with remainder summation $\sum^{\infty}_{r=R}(\alpha_r T_r)^3 = o\{1/(\log R)\}$. If $T_r$ is fixed as $T_r = T_0$, the above two conditions reduce to (i) $\sum^{\infty}_{r=1}\alpha_r= \infty$ with $\alpha_R= o\{1/(\log R)\}$, and (ii) $\sum^\infty_{r=1}\alpha^3_s <+\infty$ with remainder summation $\sum^{\infty}_{r=R}\alpha_r^3 = o\{1/(\log R)\}$. It is remarkable that condition (ii) is surprisingly different from a wide range of SGD literature, which usually requires that $\sum^\infty_{r=1}\alpha^2_r <+\infty$ \citep{robbins1951stochastic, nemirovski2009robust, gitman2019understanding}. By this condition, we find that larger diminishing learning rates are allowed by our BMGD algorithm. Therefore, faster numerical convergence rates can be achieved.

Next, we give some discussion about several commonly used learning rate strategies. For simplicity purpose, we do not discuss $\alpha_r$ and $T_r$ separately. First, we consider the polynomial decay strategy with $\alpha_rT_r \propto r^{-\gamma}$ for some $\gamma>0$. According to previous analysis, we should have $\gamma\in(1/3,1]$ so that the two critical conditions can be satisfied according to Theorem results. Another commonly used decay strategy is the exponential decay strategy with $\alpha_rT_r \propto \gamma^{-r/b}$ for some $\gamma\in(0,1)$ and $b>0$. However, this strategy can never satisfy the critical condition (i) $\sum_{r}\alpha_r T_r= \infty$. As a result, the training procedure might stop earlier before convergence. Finally, we consider the stage-wise decay strategy. Practitioners usually use a constant learning rate to train the model for a fixed number of iterations and then reduce the current learning rate. Then whether the algorithm converges is case-dependent. For example, if the learning rates used in different stages satisfy a polynomial decay with $\gamma \in (1/3,1]$, then the stage-wise decay strategy can still leads to a convergence result. Otherwise, the algorithm might not converge.

\subsubsection{Generalized linear models}

In this section, we provide a brief discussion about the popularly used generalized linear models in the statistical literature \citep{nelder1972generalized,cantoni2001robust,kulkarni2021differentially,luo2020renewable}. Assume that conditional on $X_i$, the distribution of $Y_i$ comes from the exponential distribution family. Then the loss function for $(X_i,Y_i)$ can be defined as the two-times negative log-likelihood function, which can be analytically spelt as
\beqrs
\ell(X_i,Y_i;\theta) = -\log f(X_i,Y_i;\theta,\phi) = -\phi^{-1}\{Y_iX_i^\top\theta - b(X_i^\top \theta)\} - \log a(Y_i;\phi),
\eeqrs
where $\theta\in\mathbb{R}^p$ is the regression coefficient, $\phi$ is some positive nuisance parameter, and $b(\cdot)$ is a convex function. If we omit the nuisance parameter $\phi$, the global loss function can be defined as $\mL(\theta) = -N^{-1}\sum^N_{i=1}\big\{Y_iX_i\theta - b(X_i^\top\theta)\big\}$. Its first-order and second-order derivatives with respect to $\theta$ can be calculated as $\dot{\mL}(\theta) = -N^{-1}\sum^N_{i=1}\{Y_i - \dot{b}(X_i^\top \theta)\}X_i$ and $\ddot{\mL}(\theta) = N^{-1}\sum^N_{i=1}\ddot{b}(X_i^\top \theta)X_iX^\top_i$. Then the global estimator $\wh\theta = \operatorname{\argmax}_{\theta\in\Theta}\mL(\theta)$ is the maximum likelihood estimator (MLE). According to previous discussion in Section 2.2, we know that $\sqrt{N}(\wh\theta - \theta^*) \rightarrow_d N(0, \Sigma^{-1})$, where $\Sigma^{-1} = E\{\ddot{b}(X_i^\top \theta^*)X_iX^\top_i\}$. Then it is interested to investigate the distance between the BMGD estimator and the oracle MLE. To present our analysis, we further need some extra technical assumptions.
\begin{itemize}
     \item[(A5)] Assume that $\mu \leq \lambda_{\min} [E\{b(X_i^\top \theta)X_iX^\top_i\}]\leq\lambda_{\max} [E\{b(X_i^\top \theta)X_iX^\top_i\}]\leq L$ for all $\theta \in B(\theta^*, \rho)$, where $L$ and $\mu$ are given in assumption (A3) and (A4) and $\rho>0$ is a radius parameter.
\end{itemize}
Assumption (A5) requires the expectation of the second-order derivative $\ddot{\mL}(\theta)$ to be well-behaved in a local region of the true parameter $\theta^*$. This assumption further implies the global loss function satisfying the strongly convex condition with probability tending to one as $N\to\infty$. Note that a loss function satisfying the strongly convex condition belongs to the P\L\ function class. We then have $\|\wh\theta^{(T,M)}_{r,K} - \wh\theta\|^2 \leq \mu^{-1} \big\{\mL(\wh\theta_{r,K}^{(T,M)}) - \mL^*\big\}$ holds with probability tending to one as $N\to\infty$. Take the results of diminishing learning rate as an example, we should know that $\lim_{R\to\infty}\|\wh\theta^{(T,M)}_{R,K}-\wh\theta\|^2 \to 0$ as long as (i) $\sum^{\infty}_{r=1}\alpha_r T_r= \infty$ with $\alpha_R T_R = o\{1/(\log R)\}$, and (ii) $\sum^{\infty}_{r=1}(\alpha_r T_r)^3 < +\infty$ with remainder summation $\sum^{\infty}_{r=R}(\alpha_r T_r)^3 = o\{1/(\log R)\}$ hold.

\section{Numerical studies}

\subsection{Two simulation studies}

\par In this subsection, we present several simulation studies to demonstrate the finite sample performance of the proposed BMGD method. The statistical properties of the BMGD estimator are investigated with simulated data. Both a linear regression model and logistic regression model are considered. The effects of various tuning parameters (i.e. the learning rate $\alpha$, the number of epochs within buffer $T$, the mini-batch size $n$ and the buffer number $K$) are evaluated. The study of linear regression is presented to verify the theoretical claims made in Theorem \ref{thm2}. The study about logistic regression is presented to the proposed theoretical claims made in Theorem \ref{thm3}.

We start with the simplest example: a linear regression model. We investigate the finite sample performance of the BMGD methods from two different perspectives. First, we try to demonstrate the statistical properties of the BMGD estimator compared with the global ordinary least square (OLS) estimator. Second, we aim to understand how tuning parameters (i.e. the learning rate $\alpha$, the number of epochs within buffer $T$, the mini-batch size $n$ and the buffer number $K$) influence the performance of the BMGD estimator. Specifically, consider a standard linear regression model setting with a total sample size $N = 100,000$ and predictor dimension $p = 500$. For $1\leq i\leq N$, the predictor $X_i$ is independently generated from the multivariate normal distribution with mean zero and covariance matrix $\Sigma_{xx} = (\sigma_{ij})_{p\times p}$, where $\sigma_{ij} = 0.8^{|i-j|}$ for $1\leq i,j \leq p$. The error term $\varepsilon_i$ is independently generated from $N(0,1)$. Also, we generate the regression coefficient $\theta$ from a multivariate standard normal distribution and normalize it such that $\operatorname{var}(X^\top_i\theta) = 1$. Once
$\theta$ is generated, it is then fixed throughout the rest of the simulation
. Then, the response variable $Y_i$ can be calculated as $Y_i = X^\top_i\theta + \varepsilon_i$.

\begin{figure}[ht]
	\centering
	\setlength{\abovecaptionskip}{1pt}
	\includegraphics[width=1\columnwidth]{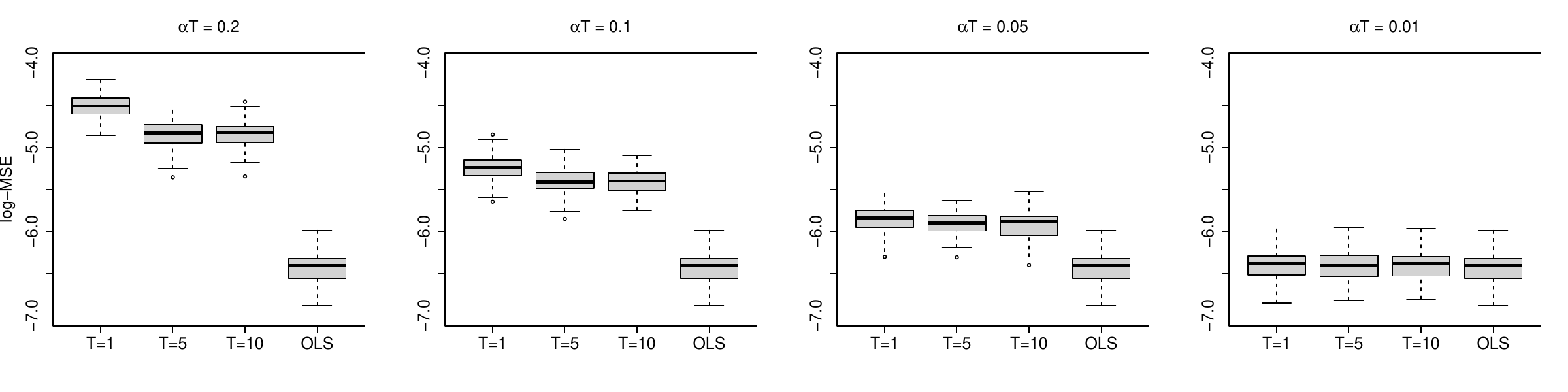}
	\caption{Boxplots of log-transformed MSE values of BMGD estimators under four different $\alpha T$ values (i.e., $\alpha T = 0.2, 0.1, 0.05, 0.01$). For a fixed $\alpha T$, three different $(\alpha,T)$ combinations are considered. The resulting log-transformed MSE values are compared with the OLS estimator.}
	\label{fig:LinearReg}
\end{figure}

We first verify the relationship between the learning rate $\alpha$ and the epoch number $T$. The buffer size is set to be $N_0 = 10,000$, and the mini-batch size is fixed at $n = 1,000$. We set $\alpha T = 0.1,0.01, 0.001$. For a fixed $\alpha T$, we consider three different $(\alpha, T)$ combinations with $T = 1, 5, 10$. The total number of iterations is fixed to be $R = 100$. The statistical efficiency of the resulting estimators is then evaluated by the mean squared error (MSE) and compared with the global OLS estimator. Specifically, it is calculated as $\widehat{\operatorname{MSE}}(\widehat\theta) = B^{-1}\sum^B_{b=1}\|\widehat\theta^{(b)} - \theta\|^2$. The entire simulation experiment is replicated a total of $B = 100$ times for each setting. The resulting MSE values are log-transformed and shown as a box plot in Figure \ref{fig:LinearReg}.

By Figure \ref{fig:LinearReg}, we can draw the following conclusions. First, the statistical properties of BMGD estimators are strongly influenced by $\alpha T$. Larger $\alpha T$ are always associated with larger MSE values, which agrees with the theoretical claims in Theorem \ref{thm2}: the distance between the BMGD estimator and the OLS estimator is linearly bounded by $\alpha T$. The BMGD estimator also becomes statistically efficient if $\alpha T$ is sufficiently small. Second, we find that the MSE values of the BMGD estimator remain stable when the number of epochs within buffer $T$ increases if $\alpha T$ is fixed. Again, this finding agrees the claims made in Theorem \ref{thm2}. If we treat the classical MGD algorithm as a special case of BMGD but with $T=1$. Then, a similar estimation accuracy can be achieved by a larger $T$ but smaller learning rate $\alpha$. The time cost is also greatly reduced, as shown later.

\begin{figure}[ht]
	\centering
	\setlength{\abovecaptionskip}{1pt}
	\includegraphics[width=1\columnwidth]{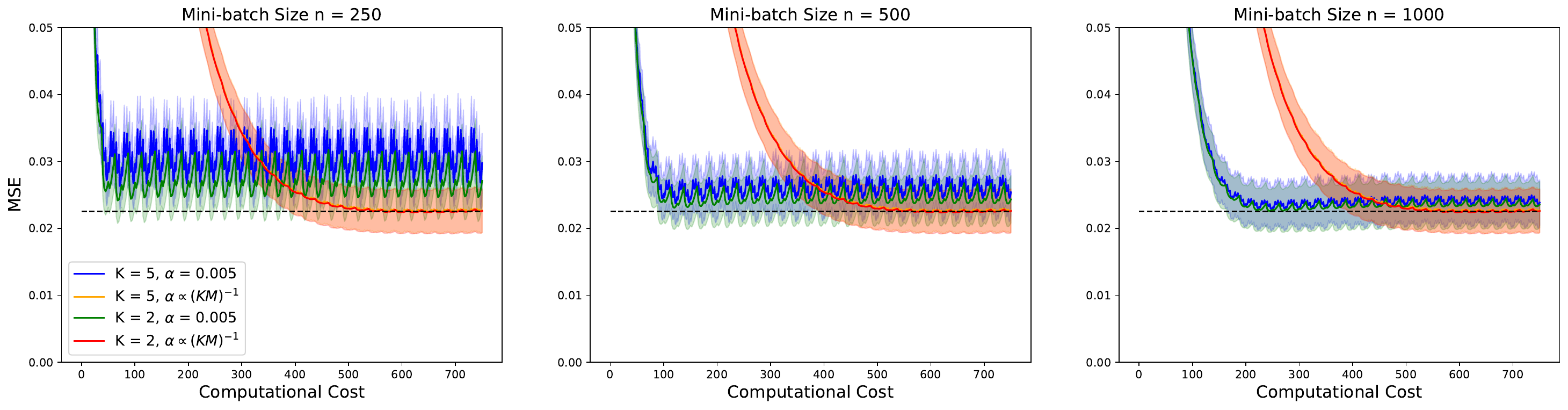}
	\caption{Plot of the mean MSE values of BMGD estimators during training process for a total of $B = 100$ replicates with $95\%$ confidence interval shades. Three different mini-batch size $n$ values (i.e., $n = 250, 500, 1000$) are considered. For a fixed mini-batch size $n$, four different $(\alpha, K)$ combinations are considered. The mean MSE value of the OLS estimator is marked as the black dashed line.}
	\label{fig:LinearBuffer}
\end{figure}

Next, we consider the relationship between the learning rate $\alpha$, the mini-batch size $n$, and the number of buffer $K$. We fix the epoch number within buffer $T$ to be 5 and the total number of iterations to be $R = 30$. We consider three different mini-batch sizes $n = 250, 500, 1,000$, two different buffer numbers $K = 2, 5$, and two different learning rates $\alpha = 0.005$ or $\alpha = 1/(TKM)$. This leads to a total of $3\times 2\times2 = 12$ different combinations. The entire simulation experiment is replicated a total of $B = 100$ times for each setting. The statistical efficiency of the resulting estimators is then evaluated by the mean squared error (MSE) and compared with the global OLS estimator. The resulting MSE values along the training process are recorded and are shown in Figure \ref{fig:LinearBuffer}.

It is remarkable that we use computational cost in Figure \ref{fig:LinearBuffer} to replace the CPU time or iteration number. Here, one unit of computational cost in Figure \ref{fig:LinearBuffer} represents the computational cost for calculating a gradients of sample size $20,000$. Consequently, we observe the performance of different BMGD estimators after the same amount of calculation. By Figure \ref{fig:LinearBuffer}, we can draw the following conclusions. First, the statistical property of the BMGD estimators are influenced by the mini-batch size $n$ and buffer number $K$. When the mini-batch size $n$ becomes smaller or the buffer number $K$ becomes larger, a smaller learning rate should be used for better statistical accuracy. Specifically, we find that using a learning rate $\alpha$ that is proportional to $1/(KM)$ leads to satisfactory performance under different mini-batch sizes and buffer numbers. This is consistent with our theoretical discussions for Theorem \ref{thm3}. Furthermore, we find that for a learning rate $\alpha$ that is proportional to $1/(KM)$, the BMGD estimators have similar performances
after the same amount of computational cost among different settings. This property gives us an opportunity to find better tuning parameter combinations to reduce the time cost. 

\begin{figure}[ht]
	\centering
	\setlength{\abovecaptionskip}{1pt}
	\includegraphics[width=1\columnwidth]{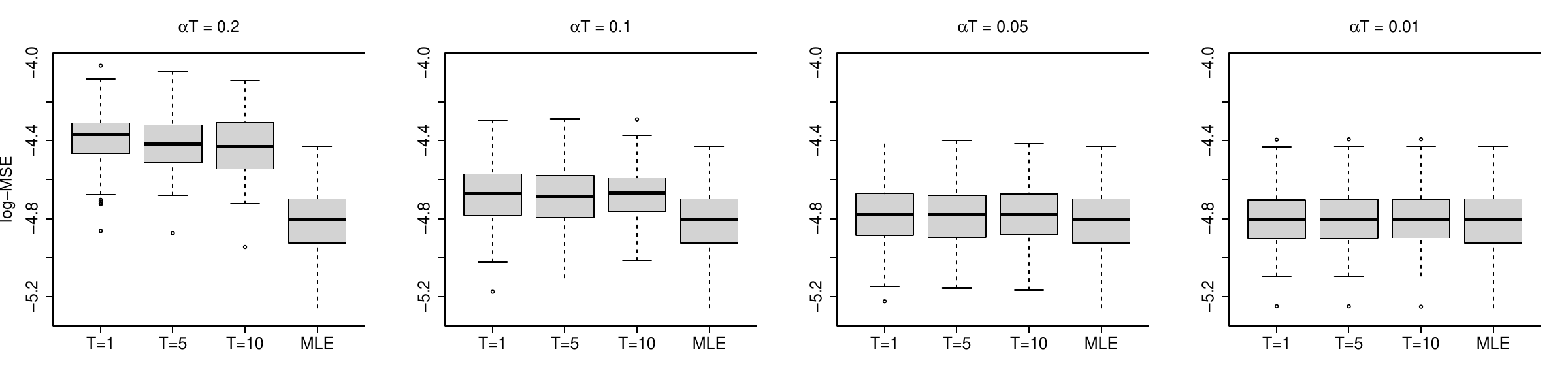}
	\caption{Boxplots of log-transformed MSE values of BMGD estimators under four different $\alpha T$ values (i.e., $\alpha T = 0.2, 0.1, 0.05, 0.01$). For a fixed $\alpha T$, three different $(\alpha,T)$ combinations are considered. The resulting log-transformed MSE values are compared with the maximum likelihood estimator (MLE).}
	\label{fig:LogisticReg}
\end{figure}

\begin{figure}[ht]
	\centering
	\setlength{\abovecaptionskip}{1pt}
	\includegraphics[width=1\columnwidth]{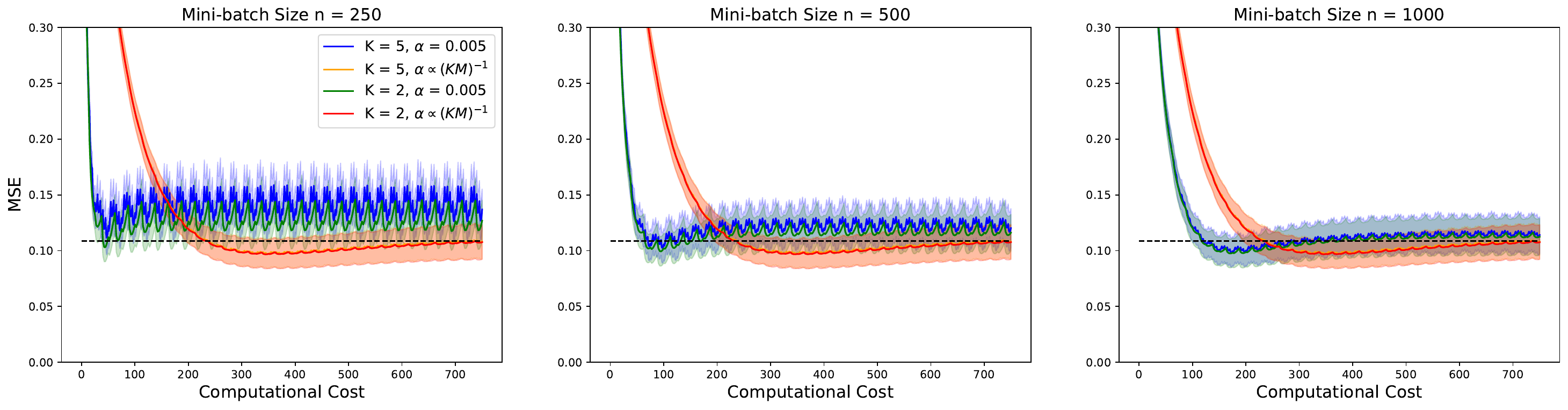}
	\caption{Plot of the mean MSE values of BMGD estimators during training process for a total of $B = 100$ replicates with $95\%$ confidence interval shades. Three different mini-batch size $n$ values (i.e., $n = 250, 500, 1000$) are considered. For a fixed mini-batch size $n$, four different $(\alpha, K)$ combinations are considered. The mean MSE value of the maximum likelihood estimator is marked as the black dashed line. 
 }
	\label{fig:LogisticBuffer}
\end{figure}

Furthermore, we also consider to demonstrate the finite sample performance of the BMGD algorithm on more general loss functions. Specifically, we consider a standard logistic regression model. Similar to the simulation experiment for linear regression, we fix the total sample size $N = 100,000$ and predictor dimension $p = 500$. The predictors $X_i$ and regression coefficient $\theta$ are generated as before. Once $X_i$ is obtained, the response variable $Y_i$ can be generated accordingly from a Bernoulli distribution with the probability given as $P(Y_i = 1|X_i, \theta) = \exp(X_i^\top \theta)/\{1 + \exp(X_i^\top \theta)\}.$ The loss function evaluated here is two times the negative log-likelihood function. By minimizing this loss function, the maximum likelihood estimator (MLE) can be computed. The experiments are randomly replicated in the same way for linear regressions. The detailed results are illustrated in Figure \ref{fig:LogisticReg} and \ref{fig:LogisticBuffer}. The key findings are qualitatively similar to those of the linear regression model.

\subsection{Tsinghua-Tencent 100K dataset}

The size of the simulated data presented in the previous subsection is relatively small so that the simulation study can be accomplished quickly. However, we thus lose the opportunity to demonstrate the computational efficiency of the proposed BMGD algorithm. To this end, more sophisticated models and datasets of massive sizes are required. In this regard, three image-related massive datasets are investigated subsequently. We start with the Tsinghua-Tencent 100K dataset with a logistic regression model. Specifically, the dataset considered here is the prohibitory traffic sign-related images contained in the Tsinghua-Tencent 100K dataset \citep{ZheZhu2016TrafficSignDA}. The dataset contains a total of 8,496 high resolution (2,048$\times$2,048) color images, including 5,662 images in the training set and 2,834 images in the testing set. This dataset can be publicly obtained at \url{http://cg.cs.tsinghua.edu.cn/traffic-sign/}. The objective of this study is to train a logistic regression model so that the traffic signs contained in the images can be detected automatically, which is an important problem for automatic driving. Thus, each image has been annotated in the sense that tight bounding boxes have been created for the traffic signs of interest.

The image resolution (i.e., 2,048$\times$2,048) of this dataset is high. Comparatively speaking, the region containing traffic signs is much smaller, which makes the intended detection problem challenging and thus calls for appropriate simplification. In this regard, we first partition the original image into 64$\times$64 subimages with each subimage of size 32$\times$32, which leads to a total of 64$\times$64=4,096 samples for each image. Because a total of 5,662 images are involved in the training set, the total sample size in this study is then $5,662\times64\times64=23,191,552$, which is large. We then treat each subimage as one sample. For each sample, a response label is created and set to be 1 if a traffic sign is contained in the subimage or 0 otherwise. Here, a subimage is considered to contain a traffic sign if more than 20\% of its area is covered by the bounding box of a traffic sign. For each subimage, we transfer a pretrained VGG16 network so that a feature vector of $p=512$ dimensions can be created \citep{KarenSimonyan2015VeryDC}. After feature generation, each image is transformed to a $64\times64\times512$ tensor and placed on the hard drive. This process requires more than 67 GB on the hard drive.

The number of positive instances is also much smaller than that of the negative instances. The number of positive instances only accounts for approximately 0.235\% of the total sample, which is expected because the region containing traffic signs in a high relation image is small. The class imbalance issue may negatively affect the training process. For a reliable training process, the negative instances are appropriately undersampled for each buffer \citep{HaiYingWang2020LogisticRF}. Thus, the numbers of positive and negative instances are approximately balanced for each mini-batch. The resulting estimators (i.e., the MGD or BMGD estimator) are then applied to this dataset. Once again, the model used here is a standard logistic regression model. Both BMGD and the traditional MGD algorithms are used to compute the maximum likelihood estimator for the unknown regression coefficient. For both algorithms, the Adam algorithm of \cite{kingma2014adam} is used to compute the gradient direction. The initial learning rate is set to 0.001. All other tuning parameters (e.g., the decay rate) are set to be the defaults of TensorFlow 2.6. Both algorithms are trained for a sufficient amount of time. The mini-batch size is set to 1,000 for both BMGD and MGD.

The resulting estimators are then applied to the test dataset so that the response probability can be computed for each test sample. We recall that for each image, a total of 64$\times$64=4,096 subimages (i.e., samples) can be created. Thus, a standard AUC \citep{CharlesXLing2003AUCAS}, i.e., the area under the receiver operating characteristic curve, can be computed at the image level. This image-level AUC is then averaged across different images, which leads to the overall performance measure $\overline{\mbox{AUC}}$. Furthermore, a commonly used evaluation criterion for classification, the F1 score, is also adopted \citep{Fmeansure}. In a similar way, we compute image-level F1 scores and then calculate the averaged metric across different images, which is denoted as $\overline{\mbox{F1}}$. Given that positive instances are sparse in the test samples, we also considered another interesting measure. Specifically, for a given image with 64$\times$64=4,096 samples, we then predict those samples with estimated response probability larger than a prespecified threshold value as positives and others as negatives. The threshold value is set to be the largest threshold value so that all positive samples can be captured. Thus, we inevitably pay a price by mistakenly predicting some negative samples as positives, which leads to a number of false-positive predictions. Their total number is then recorded as a False Positive (FP)
number, which is the number computed at the image level. This number can then be further averaged across different images, which leads to the final performance measure $\overline{\mbox{FP}}$.

\begin{figure}[ht]
	\centering
	\setlength{\abovecaptionskip}{1pt}
	\includegraphics[width=1\columnwidth]{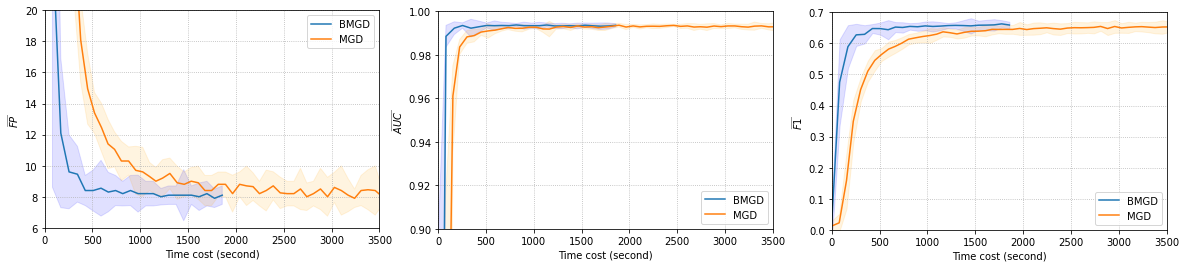}
	\caption{Optimization performances of BMGD and MGD.}
	\label{fig:ts_auc_fp}
\end{figure}

Detailed results are shown in Figure \ref{fig:ts_auc_fp}. For both the BMGD and MGD algorithms, we repeat the training process 10 times with different random initial values. The averaged results are reported in Figure \ref{fig:ts_auc_fp}. In Figure \ref{fig:ts_auc_fp}, the solid lines display the averaged results. For each solid line, the corresponding shading reflects a 95\% confidence interval across multiple runs. The left, middle, and right panels demonstrate the $\overline{\mbox{FP}}$ results, the $\overline{\mbox{AUC}}$ results and the $\overline{\mbox{F1}}$ results, respectively. Both results clearly show that the BMGD algorithm is computationally more efficient than the traditional MGD algorithm. Specifically, compared with the traditional MGD algorithm, the BMGD algorithm can achieve competing performance in terms of $\overline{\mbox{FP}}$, $\overline{\mbox{AUC}}$ and $\overline{\mbox{F1}}$ but much more quickly. For example, the $\overline{\mbox{FP}}$ value of the BMGD algorithm approached the minimum value of 8 after training for approximately 1,000 seconds. In contrast, approximately 2,700 seconds are required for the MGD algorithm for similar performance. Comparatively speaking, the BMGD algorithm saves approximately $(2,700-1,000)/2,700=62.96\%$ of the requested time. Moreover, we adopt Wilcoxon rank sum test to verify the differences between the performances of BMGD and MGD. Given a fixed training time, we compare the performances of BMGD in multiple runs with those of MGD in terms of $\overline{\mbox{FP}}$, $\overline{\mbox{AUC}}$ and $\overline{\mbox{F1}}$. If the Wilcoxon rank sum test is rejected, it is suggested that there exists significant difference between the classification performances of the two methods. Table \ref{wtest} shows the test results with the training time ranges from 200 seconds to 1,500 seconds. The results demonstrate that in the early stage of training, e.g., when the training time is no more than 500 seconds, the performance of BMGD is significantly better than MGD. Thus, we conclude that compared with MGD, the BMGD method, which incorporates a buffering mechanism, provides a more efficient solution for model training.

\begin{table*}
\bc\emph{}
\caption{\label{wtest}Wilcoxon rank sum test results with different training time. }
\begin{threeparttable}
\begin{tabular}{c|cl|cl|cl}
\hline
Training time &  \multicolumn{2}{c|}{$\overline{\mbox{FP}}$}  & \multicolumn{2}{c|}{$\overline{\mbox{AUC}}$} & \multicolumn{2}{c}{$\overline{\mbox{F1}}$}    \\
(seconds) & statistic & p-value & statistic & p-value & statistic & p-value   \\
\hline
200 & 3.7796 & 0.0002 & 3.2505 & 0.0012 & 3.7041 & 0.0002 \\
500 & 3.7418 & 0.0002 & 2.9481 & 0.0032 & 3.4773 &  0.0005 \\
1000 & 3.2505 & 0.0011 & 2.8725 & 0.0041 & 3.4017 & 0.0007\\
1500 & 2.7591 & 0.0058 & 1.8142 & 0.0696 & 3.2505 & 0.0012\\
\hline
\end{tabular}
\end{threeparttable}
\ec
\end{table*}

\begin{figure}[ht]
	\centering
	\setlength{\abovecaptionskip}{1pt}
	\includegraphics[width=1\columnwidth]{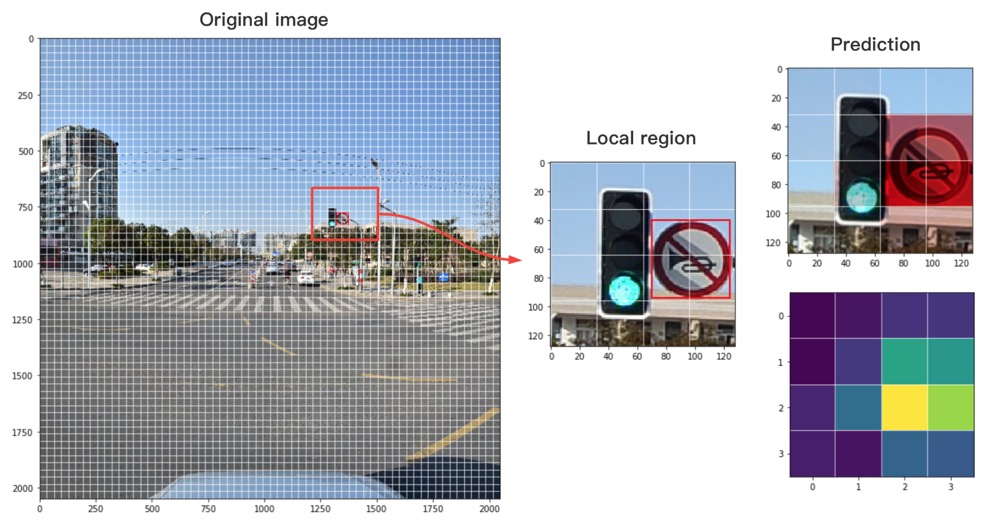}
	\caption{A case of the prediction for traffic sign detection}
	\label{fig:ts_case}
\end{figure}

Finally, to intuitively demonstrate the performance of the model trained by BMGD, we present a concrete prediction example in Figure \ref{fig:ts_case}. The left panel demonstrates a test sample of high resolution. For this image, there is an interesting traffic sign (i.e., a sign showing ``No Tooting'') sitting on the right side of the traffic light. The position is near the northeast corner of the entire image. For a better view, this local region is separately displayed in the middle panel. The predicted result is given in the right panel, which seems to be very accurate.

\subsection{CatDog dataset}

We next consider the CatDog dataset and a more sophisticated deep learning model (i.e., AlexNet) in our second case study. The original dataset was released by Kaggle in 2013. It can be obtained at \url{https://www.kaggle.com/competitions/dogs-vs-cats/data/}. The dataset contains 15,000 training images and 10,000 test images that belong to two balanced classes: cats and dogs. The entire dataset is stored on the hard drive in JPEG format and occupies approximately 600 MB of storage space in total. If we want to load all images into memory in tensor format with dimension $128\times 128 \times 3$, then more than 20 GB CPU memory is required. For better generalization performance, various data augmentation techniques have been popularly used for deep learning \citep{Shorten2019ASO}. These data augmentation techniques include shearing with range $[0, 0.5]$, rotating with maximum angle $30$, zooming with scale range $[0.8, 1.2]$, shifting for both height and width with range $[0, 0.2]$ and horizontal flipping. Thus, the size of the augmented dataset should be significantly larger than that of the original dataset, which makes computation challenging and calls for more efficient solutions. The major task of this dataset is to train a classical convolutional neural network (CNN) model for image classification. Specifically, we consider a classic CNN model, AlexNet, as our classifier. The AlexNet model was first proposed by \cite{2012AlexNet} and won the championship on ImageNet Large Scale Visual Recognition Challenge (ILSVRC) in 2012. The model takes a $128\times 128$ pixel RGB image (i.e. $128\times 128\times 3$ tensor) as input. The entire model contains a total of 19 layers, including 5 convolution layers and 3 fully connected layers, which requires 24,737,858 parameters.

The AlexNet model is then trained on the CatDog dataset with the help of a GPU-CPU system with a total of 5 nodes. Each node provides a powerful CPU and GPU. The CPU used in this study is Intel(R) Xeon(R) Gold 6271 with 12 cores and 64 GB of memory. The GPU used in this study is a NVIDIA Tesla P100 with 16 GB of memory. To efficiently execute the traditional MGD algorithm, the distributed computing strategy of TensorFlow 2.6 is used. The mini-batch size per GPU is set to be $150$, which leads to a $5\times 150 = 750$ global mini-batch size in total. Since the proposed BMGD framework mainly focuses on the data pre-loading strategy, it can be readily combined with various gradient-based optimization algorithms. For illustration purpose, we consider here three different optimization algorithms: mini-batch gradient descent (MGD), nesterov accelerated gradient (NAG) and Adam \citep{kingma2014adam}. By combining the buffering idea with these methods, the buffered optimization algorithms can be readily developed. For the baseline methods, the model is trained for a total of 80 epochs. We consider the step learning decaying strategy here. The initial learning rate is set to be $0.1$, $0.05$ and $0.001$ for MGD, NAG and Adam, respectively.  After 70 epochs the learning rate becomes one-tenth of the original one. For the buffered optimization algorithms, we set the buffer size equal to the original sample size but with different data augmentations executed for each buffer. The model is trained in 4 stages. In the first stage, the model is trained on 4 different buffers and number of buffer epoch is set to be $10$. In the second stage, the model is trained on 4 different buffers and number of buffer epoch is set to be $5$. In the third stage, the model is trained on 5 different buffers and number of buffer epoch is set to be $2$. In the last stage, the model is trained on 10 different buffers and number of buffer epoch is set to be $1$. Consequently, the total number of epochs excuted by the buffered version methods is also $80$, which leads to approximately the same amount of computational cost compared with their baseline methods. The initial value of learning rates for the three buffered version methods are the same as their baseline method, and they decay to be one-tenth of the original one in the last training stage.

\begin{figure}[ht]
	\centering
	\setlength{\abovecaptionskip}{1pt}
	\includegraphics[width=1\columnwidth]{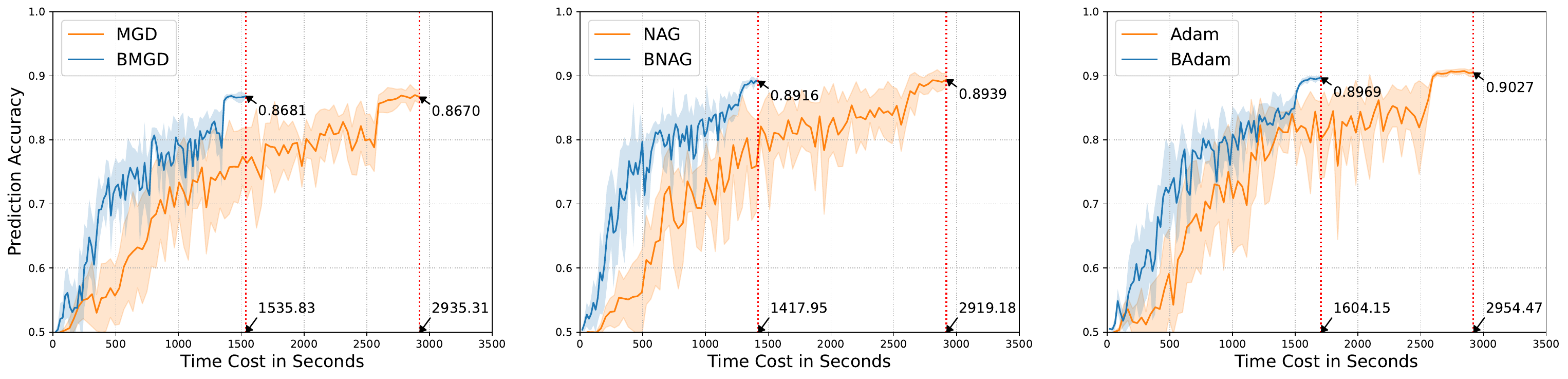}
	\caption{The prediction accuracy evaluated on the CatDog test dataset. A total of $10$ random replications are conducted. Three optimization algorithms and their buffered competitors are evaluated. The reported statistics are mean out-of-sample prediction accuracy results with $95\%$ confidence interval shades.}
	\label{fig:CatDog}
\end{figure}

We use MGD, NAG and Adam to denote the three baseline optimization algorithms. We use BMGD, BNAG and BAdam to denote the three buffered optimization algorithms. The estimators obtained by all the algorithms are evaluated on the test data in terms of prediction accuracy. The associated computation time costs are also recorded. For a reliable evaluation, the experiments are randomly replicated for a total of $10$ times. The mean of replicated results are then given in Figure \ref{fig:CatDog} with $95\%$ confidence interval shades. By Figure \ref{fig:CatDog},  we find that the time costs required by the buffered algorithms are only about $50\%$ of their baseline counterparts for achieving comparable prediction performances. Take the NAG algorithm as an example. The BNAG method can achieve about $89.16\%$ in mean prediction accuracy on the test dataset within $1417.95$ seconds. In contrast, it takes about $2919.18$ seconds for a standard NAG method to achieve about $89.39\%$ in mean prediction accuracy. This is a result slightly better than $89.16\%$ of the BNAG method. But the time cost paid by a standard NAG method is about twice that of the BNAG method.

\subsection{ImageNet dataset}

Lastly, we evaluate the proposed BMGD method on the ImageNet dataset, which is one of the most popularly used image datasets in deep learning researches. The dataset is available at \url{https://image-net.org/download.php}. It consists of 1,281,167 training images and 50,000 validation images with 1,000 object classes. All images are stored in JPEG format and take about 144 GB of storage space on the hard drive. The objective here is to train a ResNet50 model of \citet{He2016Deep} for image classification. The model contains a total of 25,593,792 parameters and it takes a $224\times 224 \times 3$ tensor as input. We adopt the same computational environment as before. That is a GPU-CPU system with a total of 5 nodes as we described in Section 3.3. The mini-batch size per GPU is set to $128$, which leads to a global batch with size 640. For the traditional MGD method, the number of epochs is set to be 50. For the proposed BMGD method, the number of buffer $K$ is set to be $10$. Consequently, the size of each buffer is $1/10$ of the size of the whole training set, which contains about $1.2\times 10^5$ images. The model is trained for $4$ stages. In the first stage, the model is trained for $2$ iterations and $T$ is set to be $5$. In the second stage, the model is trained for $3$ iterations and the number of buffer epoch $T$ is set to be $10$. In the third stage, the model is trained for $5$ iterations and $T$ is set to be $2$. In the last stage, the model is trained for $5$ iterations and $T$ is set to be $1$.

\begin{figure}[ht]
	\centering
	\setlength{\abovecaptionskip}{1pt}
	\includegraphics[width=1\columnwidth]{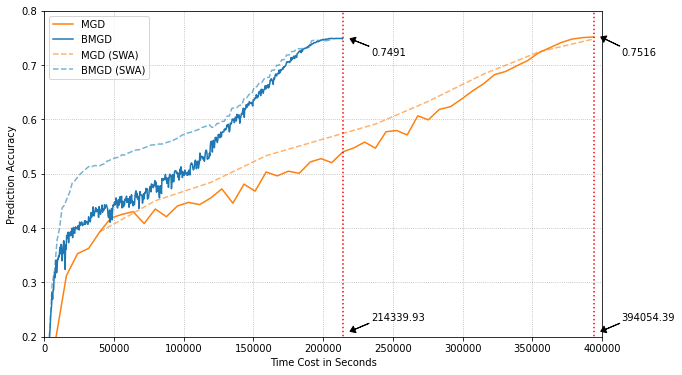}
	\caption{The prediction accuracy evaluated on the ImageNet validation dataset. The MGD algorithm, the MGD algorithm with SWA, and their buffered competitors.}
	\label{fig:Imagenet}
\end{figure}

To achieve an acceptable prediction performance for the ImageNet dataset with ResNet50, various refinement operations as suggested by \cite{he2019bag} have been implemented. Specifically, these operations are, respectively, decoding the pixel values in $[0,255]$, random cropping, horizontal flpping with 0.5 probability, scaling hue, saturation, brightness with random coefficients, adding noises based on principal component analysis for the training dataset, and normalizing RGB channels. Those operations are useful for improving the generalization performance on the validation dataset. Nevertheless, the price paid here is a significant increase of time cost for data processing. Lastly, an SGD optimizer with cosine learning rate decay as well as learning rate warmup is used \citep{he2019bag}. The initial value for the learning rate is set to be 0.05. Then, the learning rate for the $j$-th update is set to be $\alpha_{r,j} = 0.025+0.025\cos(j\pi/TMK)$ with $1\leq j\leq TMK$ for every iteration $r$. We utilize the prediction accuracy on the validation set as the evaluation metric. Figure \ref{fig:Imagenet} displays the performances for MGD and BMGD in terms of prediction accuracy. The result shows that the proposed BMGD method is computationally more efficient than the traditional MGD method. Specifically, BMGD achieves a 74.91\% prediction accuracy on the validation set within about 60 hours. While for MGD, it takes about 110 hours, which is nearly 1.84 times the cost of BMGD, to achieve a similar level of prediction performance. 

Moreover, since the proposed BMGD framework mainly focuses on the data pre-loading strategy, it can be readily combined with various gradient-based algorithm and training strategies. For illustration purpose, we consider here the stochastic gradient weight averaging method (SWA) proposed by \citet{SWA2018}. The SWA method is a powerful ensemble learning strategy designed for the deep neural networks to achieve better generalization performance. It averages the model weights after a pre-specified number of epochs. The averaged weight could lead to a better generalization performance. By combining our buffering idea with the SWA method, a buffered SWA method can be readily developed. For the MGD algorithm, the SWA procedure is set to calculate the moving average of the model weights after every 5 epochs. For the proposed BMGD algorithm, the SWA procedures is set to calculate the moving average of the model weights after every 5 buffer epochs. In Figure \ref{fig:Imagenet}, the blue dashed line and orange dashed line display the performances of SWA method for the BMGD algorithm and MGD algorithm, respectively. We find that given the same training time, the performance of SWA based on BMGD is better than that based on MGD. This suggests that for SWA, the incorporation of BMGD can further improve its training performance.

\section{Concluding remarks}

In this paper, we study a buffered mini-batch gradient descent (BMGD) algorithm for training sophisticated model on massive datasets. The proposed algorithm is designed for fast training on a GPU-CPU system, which contains two primary steps: the buffering step and the computation step. In the buffering step, a large amount of data (i.e., a buffer) are pre-loaded into the system RAM of CPU, and then loaded into the graphical memory of GPU. In the computation step, a standard MGD algorithm is applied to the buffered data.

Compared to the traditional MGD algorithm, the proposed BMGD algorithm can be much more efficient for two reasons. First, a mini-batch of data is used to compute the gradient update only once when it is loaded into the system RAM of CPU in the MGD algorithm, while the BMGD algorithm uses the buffered data for multiple epochs. Given that the time cost due to communication from the hard drive to the system RAM of CPU and the graphical memory of GPU is quite expensive, the communication cost of the proposed BMGD algorithm can be remarkably reduced. Second, the buffering step can be executed in parallel so that the GPU does not remain idle when loading new data. Therefore, the total time cost can be further reduced.

We comprehensively investigate the theoretical properties of BMGD algorithms. We start with the linear regression model and the squared loss function. We find that under appropriate conditions, the BMGD estimator should converge to a stable solution. Then, the asymptotic properties of the stable solution are investigated under a constant learning rate $\alpha$ and epoch number $T$. With a sufficiently small $\alpha T$, the BMGD estimators can be statistically as efficient as the global OLS estimator. Furthermore, we extend the theoretical analysis to a more general loss function class (i.e., the P\L- function class). We rigorously show the convergence rate of the BMGD algorithm and discuss the tuning parameter effect. The theoretical claims are then shown via extensive simulation studies. To demonstrate the marked advantage in computational efficiency of the proposed method, three real image-related datasets are investigated, and more sophisticated deep learning models (i.e. CNN) are considered.

To summarize this article, we discuss several interesting directions of future research. First, although the P\L-function classes includes many non-convex functions, it is still not large enough to include many deep learning models as its special cases. It is then important to investigate the asymptotic properties of the BMGD algorithm under a more general loss function family. 
Second, although the theoretical results presented in this paper are restricted to the constant learning rate. We find various learning rate decaying strategies work well in the real data experiments. Therefore, it is also important to provide theoretical guidance for those learning rate decaying strategies for real practice.


\newpage
\bibliographystyle{plainnat}
\bibliography{ref}

\end{document}